\documentclass[12pt]{article}
\usepackage[utf8]{inputenc} 
\usepackage[english]{babel} 
\usepackage{lipsum} 
\usepackage{url} 
\usepackage[dvipsnames]{xcolor}
\usepackage{authblk}
\usepackage{tcolorbox}
\usepackage{float}
\usepackage{mathtools}
\usepackage{booktabs} 
\usepackage{fancyhdr} 
\usepackage{hyperref} 
\usepackage{amsmath} 
\usepackage[toc,page]{appendix}
\usepackage{bbold}
\usepackage{blindtext}
\usepackage{caption}
\usepackage{subcaption}
\usepackage{amsthm} 
\usepackage{mathtools} 
\usepackage{amsthm} 
\usepackage{bm} 
\usepackage{braket} 
\usepackage{tikz} 
\usepackage{amssymb} 
\usepackage{amsfonts} 
\newcommand{\R}{\mathbb{R}}

\newcommand{\Z}{\mathbb{Z}}
\newcommand{\de}{\partial}
\newcommand{\diff}{\mathrm{d}}

\newcommand{\ds}{\diff s}

\newcommand{\be}{
\begin{equation}
}

\newcommand{\beLabel}[1]{
\begin{equation}
\label{eq:#1}
}

\newcommand{\ee}{
\end{equation}
}

\newcommand{\eeLabel}{
\end{equation}
}

\let\oldref\ref
\renewcommand{\ref}[1]{(\oldref{#1})}
\title{\textbf{
\vskip-3cm
Geometric Flow of Bubbles}}
\date{}
\author[a]{Davide De Biasio}
\author[a,b]{Dieter L\"ust}
\affil[a]{Max--Planck--Institut f\"ur Physik, Werner--Heisenberg--Institut, \newline
F\"ohringer Ring 6, 80805 M\"unchen, Germany}
\affil[b]{Arnold Sommerfeld Center for Theoretical Physics, \newline
Ludwig Maximilians Universit\"at M\"unchen, \newline Theresienstrasse 37, 80333 M\"unchen, Germany}

\usepackage{cite}
\numberwithin{equation}{section}
\begin{document}
\fancypagestyle{plain}{%
	\fancyhead[R]{LMU-ASC 02/22 \\
MPP-2022-2}
	\renewcommand{\headrulewidth}{0pt}
}
\maketitle
\begin{center}Abstract:
\end{center}
In this work we derive a class of geometric flow equations for metric-scalar systems. Thereafter, we construct them from some general string frame action by performing volume-preserving fields variations and writing down the associated gradient flow equations. Then, we consider some specific realisations of the above procedure, applying the flow equations to non-trivial scalar bubble and metric bubble solutions, studying the subsequent flow behaviour. 
\newpage
\tableofcontents
\newpage
\section{Introduction}
Given the \textit{moduli space} of apparently consistent models below a given energy cut-off $\Lambda$, the \textit{Swampland Program} \cite{Vafa:2005ui,Palti:2019pca,vanBeest:2021lhn,Brennan:2017rbf} aims at pointing out decisive attributes of quantum field theories coupled to gravity, discerning those that admit an ultraviolet completion to \textit{quantum gravity} from the others. The former ones are generally said to belong to the \textit{quantum gravity landscape}, while the latter are arranged into the \textit{quantum gravity swampland}. Speculative statements regarding the classification of low energy effective theories into the ones contained in the swampland and those falling into the landscape are commonly named \textit{swampland conjectures}. In most occurrences of the above-mentioned discussions, the specific quantum gravity picture one has in mind is the one emerging from \textit{superstring} theory. Nevertheless, evidence backing up swampland conjectures is often grounded in more general considerations on quantum field theory in curved space-time and on the expected behaviour of the quantised gravitational field. Therefore, we will refer to the ultraviolet complete theory from which landscape models are expected to descend with the vague term \textit{quantum gravity} whenever possible.
The space of seemingly consistent theories below a given energy is commonly described as a geometric object, charted by a set of \textit{generalised moduli}. The \textit{Distance Conjecture}, which first appeared in \cite{Ooguri:2006in}, states that large distances in moduli space should be accompanied by infinite towers of asymptotically massless new states. Drawing inspiration from the graviton renormalization group flow in string theory $\sigma$-models, the distance conjecture was recently specified and rephrased  \cite{Kehagias:2019akr,Bykov:2020llx,Luben:2020wix,DeBiasio:2020xkv,debiasio2021geometric} in terms of \textit{geometric flow equations}, such as Hamilton's \textit{Ricci flow} \cite{hamilton1982,Perelman:2006un,chow2004ricci}
\begin{equation}
\begin{cases}
    \diff_{\lambda}g_{\mu\nu}\left(\lambda\right)=-2R_{\mu\nu}\left(\lambda\right)\ ,\\
   g_{\mu\nu}\left(0\right)=\Bar{g}_{\mu\nu}\ ,
\end{cases}
\end{equation}
in which $\lambda$ is a real flow parameter, $g_{\mu\nu}\left(\lambda\right)$ a one-parameter family of space-time metric tensors, $R_{\mu\nu}\left(\lambda\right)$ the associated Ricci scalar curvature and $\Bar{g}_{\mu\nu}$ an initial condition for the flow.
Considering the generalised moduli space accounting for the metric itself as a modulus of the effective theory and providing it with an appropriate notion of \textit{distance}, it was observed that Ricci flat fixed points of the flow usually sit infinitely far from the initial condition. Hence, moving to a more general notion of \textit{geometric flow}, it was hypothesised that any of its fixed points must be accompanied by an infinite tower of massless states in quantum gravity. In this work, we will follow the procedure with which Perelman derived his combined \textit{metric-dilaton} flow equations from an entropy functional $\mathcal{F}_{0}$ in a more general setting. Namely, we will also introduce polynomial self-interaction terms for the scalar in the entropy functional and write down a system of flow equations accounting for such deformations. Then, we will apply the machinery to the string-frame version of some metric-scalar action, deriving geometric flow equations from the dynamics of the system. Moreover, we will study the examples of a scalar \textit{bubble} solution, embedded in various space-time backgrounds, of a cosmological constant bubble in Anti de Sitter and of a simple solution of a specific metric-scalar set of equations of motion, discussing their flow behaviour under different geometric flow equations.
\newpage
\section{General Flow equations}\label{gfaaa}
In the following discussion, we derive the geometric flow equations associated to the \textit{euclidean} entropy functional 
\begin{equation}\label{funk}
        \mathcal{F}_{\left(\alpha,\beta,\gamma\right)}\left(g,\phi\right)=\int\diff^{D}x\sqrt{g}e^{-\phi}\left[R+\alpha\left(\nabla\phi\right)^2+\gamma\Delta\phi+\sum_{n=0}^{+\infty}\beta_{n}\phi^n\right]\ .
\end{equation}
for a rank-$2$ symmetric tensor $g_{\mu\nu}$ and a scalar $\phi$, where the overall constant in front of $\mathcal{F}$, which can always be reabsorbed in the flow parameter, was set to $1$. The standard metric-dilaton flow equations \cite{Kehagias:2019akr}
\begin{equation}
        \begin{cases}
    \diff_{\lambda}g_{\mu\nu}=-2R_{\mu\nu}\ ,\\
    \diff_{\lambda}f=-R-\Delta f+\left(\nabla f\right)^2\ ,\\
    g_{\mu\nu}\left(0\right)=\Bar{g}_{\mu\nu}\ ,\\
    f\left(0\right)=\Bar{f}\ ,
    \end{cases}
\end{equation}
correspond to the $\left(\alpha,\gamma,\beta_{0}\dots\right)=\left(0,1,0,0\dots\right)$ case. From this point on, we define:
\begin{equation}
\Bar{\mu}\equiv\left(\alpha,\gamma,\beta_{0}\dots\right)\ .
\end{equation}
It must be noted that the $\alpha$ parameter measures the significance of domain walls for $\phi$, as the term usually included in the \textit{Ginzburg-Landau} theory free energy functional. 
In order to derive the flow equations, we perform variations in $\phi$ and $g_{\mu\nu}$ so that:
\begin{equation}
    \delta\left(\sqrt{g}e^{-\phi}\right)=0\ .
\end{equation}
By taking the first order variation
\begin{equation}
    \begin{split}
        g_{\mu\nu}&\longmapsto g_{\mu\nu}+v_{\mu\nu}\\
        g^{\mu\nu}&\longmapsto g^{\mu\nu}-v^{\mu\nu}\\
        \phi&\longmapsto\phi+h\ ,
    \end{split}
\end{equation}
we have the following
\begin{equation}
\begin{split}
    \delta\left(\nabla\phi\right)^2&=\nabla^{\mu}\phi\nabla^{\nu}\phi v_{\mu\nu}+2g_{\mu\nu}\nabla^{\mu}\phi\nabla^{\nu}h\\
    \delta\nabla^{2}\phi&=-v^{\mu\nu}\nabla_{\mu}\nabla_{\nu}\phi-\delta\Gamma^{\lambda}_{\mu\nu}\nabla_{\lambda}\phi+\nabla^{2}h=\\
    &=-v^{\mu\nu}\nabla_{\mu}\nabla_{\nu}\phi-\frac{1}{2}\left(\nabla_{\mu}v_{\nu\xi}+\nabla_{\nu}v_{\mu\xi}-\nabla_{\xi}v_{\nu\mu}\right)\nabla^{\xi}\phi+\nabla^{2}h\\
    \delta R&=-\nabla^{2}v+\nabla^{\mu}\nabla^{\nu}v_{\mu\nu}-R_{\mu\nu} v^{\mu\nu}\\
    \delta\phi^k&=k\phi^{k-1}h\ .
\end{split}
\end{equation}
where $v\equiv g^{\mu\nu}v_{\mu\nu}$.
At this point, we analyse all the functional terms separately, singling the fields variations out. Starting from the polynomial term, we simply have:
\begin{equation}
    \delta\int\diff^{D}x\sqrt{g}e^{-\phi}\sum_{n=0}^{+\infty}\beta_{n}\phi^n=\int\diff^{D}x\sqrt{g}e^{-\phi}\sum_{n=1}^{+\infty}n\beta_{n}\phi^{n-1} h\ .
\end{equation}
Studying the kinetic term for the scalar and getting rid of boundary terms, we obtain:
\begin{equation}
\begin{split}
    \delta\int\diff^{D}x\sqrt{g}e^{-\phi}\alpha\left(\nabla\phi\right)^2=&-\int\diff^{D}x\sqrt{g}e^{-\phi}\alpha\nabla_{\mu}\phi\nabla_{\nu}\phi v^{\mu\nu}+\\
    &+\int\diff^{D}x\sqrt{g}e^{-\phi}2\alpha \nabla_{\mu}\phi\nabla^{\mu}h=\\
    =&-\int\diff^{D}x\sqrt{g}e^{-\phi}\alpha\nabla_{\mu}\phi\nabla_{\nu}\phi v^{\mu\nu}+\\
    &+\int\diff^{D}x\sqrt{g}e^{-\phi}2\alpha \left[\left(\nabla\phi\right)^2-\nabla^{2}\phi\right]h
\end{split}
\end{equation}
Now, we move to the curvature term and, removing boundary terms, get:
\begin{equation}
    \begin{split}
        \delta\int\diff^{D}x\sqrt{g}e^{-\phi}R=&-\int\diff^{D}x\sqrt{g}e^{-\phi}R_{\mu\nu}v^{\mu\nu}-\int\diff^{D}x\sqrt{g}e^{-\phi}\nabla^{2}v+\\
        &+\int\diff^{D}x\sqrt{g}e^{-\phi}\nabla^{\mu}\nabla^{\nu}v_{\mu\nu}=\\
        =&-\int\diff^{D}x\sqrt{g}e^{-\phi}R_{\mu\nu}v^{\mu\nu}-\int\diff^{D}x\sqrt{g}\nabla^{2}e^{-\phi}v+\\
        &+\int\diff^{D}x\sqrt{g}\nabla^{\nu}\nabla^{\mu}e^{-\phi}v_{\mu\nu}\ .
    \end{split}
\end{equation}
Before going on, we observe that:
\begin{equation}
    \nabla^{\nu}\nabla^{\mu}e^{-\phi}=e^{-\phi}\left(\nabla^{\nu}\phi\nabla^{\mu}\phi-\nabla^{\nu}\nabla^{\mu}\phi\right)\ .
\end{equation}
Hence, we are left with:
\begin{equation}
    \begin{split}
        \delta\int\diff^{D}x\sqrt{g}e^{-\phi}R=&-\int\diff^{D}x\sqrt{g}e^{-\phi}R_{\mu\nu}v^{\mu\nu}+\\
        &+\int\diff^{D}x\sqrt{g}e^{-\phi}\left[\nabla^{2}\phi-\left(\nabla\phi\right)^2\right]v+\\
        &+\int\diff^{D}x\sqrt{g}e^{-\phi}\left(\nabla^{\nu}\phi\nabla^{\mu}\phi-\nabla^{\nu}\nabla^{\mu}\phi\right)v_{\mu\nu}\ .
    \end{split}
\end{equation}
At this point, we study the two derivatives term:
\begin{equation}
    \begin{split}
        \delta\int\diff^{D}x\sqrt{g}e^{-\phi}\nabla^{2}\phi=&-\int\diff^{D}x\sqrt{g}e^{-\phi}v^{\mu\nu}\nabla_{\mu}\nabla_{\nu}\phi+\int\diff^{D}x\sqrt{g}e^{-\phi}\nabla^{2}h+\\
        &-\int\diff^{D}x\sqrt{g}e^{-\phi}\frac{1}{2}\left(\nabla_{\mu}v_{\nu\xi}+\nabla_{\nu}v_{\mu\xi}-\nabla_{\xi}v_{\nu\mu}\right)\nabla^{\xi}\phi=\\
        =&-\int\diff^{D}x\sqrt{g}e^{-\phi}v^{\mu\nu}\nabla_{\mu}\nabla_{\nu}\phi+\\
        &+\int\diff^{D}x\sqrt{g}e^{-\phi}\left[\left(\nabla\phi\right)^{2}-\nabla^{2}\phi\right]h+\\
        &-\int\diff^{D}x\sqrt{g}e^{-\phi}\frac{1}{2}g^{\mu\nu}\left(\nabla_{\mu}v_{\nu\xi}+\nabla_{\nu}v_{\mu\xi}-\nabla_{\xi}v_{\nu\mu}\right)\nabla^{\xi}\phi\ .
    \end{split}
\end{equation}
Focusing on the last term, we have objects like
\begin{equation}
    \begin{split}
       \frac{1}{2}\int\diff^{D}x\sqrt{g}e^{-\phi}g^{\mu\nu}\nabla^{\xi}\phi\nabla_{\alpha}v_{\rho\eta}=&\frac{1}{2}\int\diff^{D}x\sqrt{g}g^{\mu\nu}\nabla_{\alpha}\left(e^{-\phi}\nabla^{\xi}\phi\right)v_{\rho\eta}=\\
       =&\frac{1}{2}\int\diff^{D}x\sqrt{g}e^{-\phi}g^{\mu\nu}\left(\nabla_{\alpha}\nabla^{\xi}\phi-\nabla_{\alpha}\phi\nabla^{\xi}\phi\right)v_{\rho\eta}
    \end{split}
\end{equation}
with different indices. Putting them together, we obtain:
\begin{equation}
    \begin{split}
        \delta\int\diff^{D}x\sqrt{g}e^{-\phi}\nabla^{2}\phi=&-\int\diff^{D}x\sqrt{g}e^{-\phi}v^{\mu\nu}\nabla_{\mu}\nabla_{\nu}\phi+\\
        &+\int\diff^{D}x\sqrt{g}e^{-\phi}\left[\left(\nabla\phi\right)^{2}-\nabla^{2}\phi\right]h+\\
        &+\int\diff^{D}x\sqrt{g}e^{-\phi}\frac{1}{2}\left[\nabla^{2}\phi-\left(\nabla\phi\right)^{2}\right]v+\\
        &-\int\diff^{D}x\sqrt{g}e^{-\phi}\left(\nabla_{\mu}\nabla_{\nu}\phi-\nabla_{\mu}\phi\nabla_{\nu}\phi\right)v^{\mu\nu}=\\
        =&\int\diff^{D}x\sqrt{g}e^{-\phi}\left[\left(\nabla\phi\right)^{2}-\nabla^{2}\phi\right]h+\\
        &+\int\diff^{D}x\sqrt{g}e^{-\phi}\frac{1}{2}\left[\nabla^{2}\phi-\left(\nabla\phi\right)^{2}\right]v+\\
        &-\int\diff^{D}x\sqrt{g}e^{-\phi}\left(2\nabla_{\mu}\nabla_{\nu}\phi-\nabla_{\mu}\phi\nabla_{\nu}\phi\right)v^{\mu\nu}
    \end{split}
\end{equation}
Now, we consider the variation of the functional and collect all the above terms. We have
\begin{equation}
    \begin{split}
        \delta\mathcal{F}_{\left(\alpha,\beta,\gamma\right)}\left(g,\phi\right)=&\int\diff^{D}x\sqrt{g}e^{-\phi}\biggl\{2\biggl[\left(\alpha+\gamma\right)\left(\nabla\phi\right)^2+\sum_{n=1}^{+\infty}\frac{n}{2}\beta_{n}\phi^{n-1}-\left(\alpha+\gamma\right)\nabla^{2}\phi\biggr]h+\\
        &+\biggl[-R_{\mu\nu}+\left(1-\alpha+\gamma\right)\nabla_{\mu}\phi\nabla_{\nu}\phi-\left(1+2\gamma\right)\nabla_{\nu}\nabla_{\mu}\phi+\\
        &+\left(1+\frac{\gamma}{2}\right)g_{\mu\nu}\nabla^{2}\phi-\left(1+\frac{\gamma}{2}\right)g_{\mu\nu}\left(\nabla\phi\right)^2\biggr]v^{\mu\nu}\biggr\}\ .
    \end{split}
\end{equation}
Furthermore, we have
\begin{equation}\label{condition}
    \delta\left(\sqrt{g}e^{-\phi}\right)=0\Longrightarrow h=\frac{v}{2}\ .
\end{equation}
By plugging this result into the above expression, we get:
\begin{equation}
    \begin{split}
        \delta\mathcal{F}_{\left(\alpha,\beta\right)}\left(g,\phi\right)=&\int\diff^{D}x\sqrt{g}e^{-\phi}\biggl[-R_{\mu\nu}+\left(1-\alpha+\gamma\right)\nabla_{\mu}\phi\nabla_{\nu}\phi-\left(1+2\gamma\right)\nabla_{\nu}\nabla_{\mu}\phi+\\
        &+g_{\mu\nu}\sum_{n=1}^{+\infty}\frac{n}{2}\beta_{n}\phi^{n-1}+\left(1-\alpha-\frac{\gamma}{2}\right)g_{\mu\nu}\nabla^{2}\phi-\left(1-\alpha-\frac{\gamma}{2}\right)g_{\mu\nu}\left(\nabla\phi\right)^2\biggr]v^{\mu\nu}\ .
    \end{split}
\end{equation}
Hence, we introduce a flow parameter $\lambda$ and set the flow equation for the metric to be:
\begin{equation}
    \begin{split}
        \frac{\de g_{\mu\nu}}{\de\lambda}=&-2R_{\mu\nu}+g_{\mu\nu}\sum_{n=1}^{+\infty}n\beta_{n}\phi^{n-1}+2\left(1-\alpha+\gamma\right)\nabla_{\mu}\phi\nabla_{\nu}\phi-2\left(1+2\gamma\right)\nabla_{\nu}\nabla_{\mu}\phi+\\
        &+2\left(1-\alpha-\frac{\gamma}{2}\right)g_{\mu\nu}\nabla^{2}\phi-2\left(1-\alpha-\frac{\gamma}{2}\right)g_{\mu\nu}\left(\nabla\phi\right)^2\ .
    \end{split}
\end{equation}
From the condition derived in \eqref{condition}, we also get:
\begin{equation}
    \begin{split}
        \frac{\de\phi}{\de\lambda}=&-R+D\sum_{n=1}^{+\infty}\frac{n}{2}\beta_{n}\phi^{n-1}+\left[\left(1-\alpha\right)\left(1-D\right)+\gamma\left(1+\frac{D}{2}\right)\right]\left(\nabla\phi\right)^{2}+\\
        &-\left[1-D\left(1-\alpha\right)+\gamma\left(2+\frac{D}{2}\right)\right]\nabla^{2}\phi\ .
    \end{split}
\end{equation}
It can be straightforwardly observed that, by taking $\Bar{\mu}=\left(0,1,0,0\dots\right)$, we can get back to Perelman's metric-dilaton flow equations. By also introducing DeTurk's diffeomorphism, as the flow induced by some vector field $\xi^{\mu}$, we finally get the general $D$-dimensional gradient flow equations associated to the entropy functional \eqref{funk}:
\begin{equation}\label{geneq}
    \begin{split}
        \frac{\de g_{\mu\nu}}{\de\lambda}=&-2R_{\mu\nu}+g_{\mu\nu}\sum_{n=1}^{+\infty}n\beta_{n}\phi^{n-1}+2\left(1-\alpha+\gamma\right)\nabla_{\mu}\phi\nabla_{\nu}\phi-2\left(1+2\gamma\right)\nabla_{\nu}\nabla_{\mu}\phi+\\
        &+2\left(1-\alpha-\frac{\gamma}{2}\right)g_{\mu\nu}\nabla^{2}\phi-2\left(1-\alpha-\frac{\gamma}{2}\right)g_{\mu\nu}\left(\nabla\phi\right)^2+\mathcal{L}_{\xi}g_{\mu\nu}\ ,\\
        \frac{\de\phi}{\de\lambda}=&-R+D\sum_{n=1}^{+\infty}\frac{n}{2}\beta_{n}\phi^{n-1}+\left[\left(1-\alpha\right)\left(1-D\right)+\gamma\left(1+\frac{D}{2}\right)\right]\left(\nabla\phi\right)^{2}+\\
        &-\left[1-D\left(1-\alpha\right)+\gamma\left(2+\frac{D}{2}\right)\right]\nabla^{2}\phi+\mathcal{L}_{\xi}\phi\ .
    \end{split}
\end{equation}
\subsection{Geometric Flow from an Action}\label{gfa}
In the following discussion, it is shown how geometric flow equations can be straightforwardly induced from the action
\begin{equation}\label{aact}
    S=\int\diff^{D}x\sqrt{\tilde{g}}\left[\tilde{R}+\frac{1}{2}\left(\nabla f\right)^{2}-\sum_{n=0}^{+\infty}\frac{g_{n}}{n!}f^{n}\right]\ .
\end{equation}
of a $D$-dimensional metric-scalar $\left(\tilde{g},f\right)$ system. In order to do so, we construct an appropriate entropy functional of the form \eqref{funk} by rescaling both $\tilde{g}_{\mu\nu}$ and $f$, so that the volume element takes the correct
\begin{equation}
    \sqrt{g}e^{-\phi}
\end{equation}
form and the general expression \eqref{geneq} for the flow equations can be straightforwardly applied, with $\phi$ being the rescaled scalar. Furthermore, in order to make connection to Perelman's entropy functional, we exploit our rescaling so that the $\alpha$ prefactor in front of $\left(\nabla \phi\right)^2$ is just one. It must be stressed that $g_{0}$ accounts for the presence of a cosmological constant, but was incorporated into the sum for simplicity. Thus, we define:
\begin{equation}
    \tilde{g}_{\mu\nu}\equiv e^{2\varphi}g_{\mu\nu},\quad f\equiv\gamma\phi\ .
\end{equation}
By plugging these definitions into the metric, we have:
\begin{equation}
\begin{split}
     S=\int\diff^{D}x\sqrt{g}e^{\left(D-2\right)\varphi}\biggl[&R-2(D-1)\Delta\varphi-(D-2)(D-1)\left(\nabla\varphi\right)^2+\\
     &+\frac{\gamma^{2}}{2}\left(\nabla \phi\right)^{2}-e^{2\varphi}\sum_{n=0}^{+\infty}\frac{\gamma^{n}g_{n}}{n!}\phi^{n}\biggr]\ .
\end{split}
\end{equation}
For the volume element to take the appropriate form, we take:
\begin{equation}
    \varphi\equiv\frac{1}{2-D}\phi\ .
\end{equation}
Therefore, we obtain:
\begin{equation}
\begin{split}
     S=\int\diff^{D}x\sqrt{g}e^{-\phi}\biggl[&R+2\frac{D-1}{D-2}\Delta \phi+\left(\frac{\gamma^{2}}{2}-\frac{D-1}{D-2}\right)\left(\nabla \phi\right)^2+\\
     &-\exp{\left(\frac{2\phi}{2-D}\right)}\sum_{n=0}^{+\infty}\frac{\gamma^{n}g_{n}}{n!}\phi^{n}\biggr]\ .
\end{split}
\end{equation}
In order to connect ourselves to Perelman's entropy functional, we take:
\begin{equation}
    \frac{\gamma^{2}}{2}-\frac{D-1}{D-2}=1\Longrightarrow\gamma=\sqrt{\frac{4D-6}{D-2}}\ .
\end{equation}
Hence, we are left with:
\begin{equation}
    \begin{split}
     S=\int\diff^{D}x\sqrt{g}e^{-\phi}\biggl[&R+2\frac{D-1}{D-2}\Delta \phi+\left(\nabla \phi\right)^2+\\
     &-\exp{\left(\frac{2\phi}{2-D}\right)}\sum_{n=0}^{+\infty}\frac{g_{n}}{n!}\left(\frac{4D-6}{D-2}\right)^{n/2}\phi^{n}\biggr]\ .
\end{split}
\end{equation}
The last term still needs to be expressed in a more treatable way. Thus,we define
\begin{equation}
    p\equiv\frac{2}{2-D},\quad q_{n}\equiv\frac{g_{n}}{n!}\left(\frac{4D-6}{D-2}\right)^{n/2}
\end{equation}
and observe that:
\begin{equation}
    \begin{split}
        \exp{\left(p\phi\right)}\sum_{n=0}^{+\infty}q_{n}\phi^{n}=&\sum_{k=0}^{+\infty}\frac{p^{k}}{k!}\phi^{k}\cdot\sum_{n=0}^{+\infty}q_{n}\phi^{n}\ .
    \end{split}
\end{equation}
By defining
\begin{equation}
    p_{k}\equiv\frac{p^{k}}{k!}\ ,
\end{equation}
we further have:
\begin{equation}
    \begin{split}
        \exp{\left(p\phi\right)}\sum_{n=0}^{+\infty}q_{n}\phi^{n}=&\sum_{k=0}^{+\infty}p_{k}\phi^{k}\cdot\sum_{n=0}^{+\infty}q_{n}\phi^{n}=\\
        =&\sum_{k=0}^{+\infty}\sum_{n=0}^{k}p_{k-n}\phi^{k-n}q_{n}\phi^{n}=\\
        =&\sum_{k=0}^{+\infty}\phi^{k}\sum_{n=0}^{k}p_{k-n}q_{n}=\\
        =&\sum_{k=0}^{+\infty}\phi^{k}\sum_{n=0}^{k}\frac{g_{n}}{\left(k-n\right)!n!}\left(\frac{2}{2-D}\right)^{k-n}\left(\frac{4D-6}{D-2}\right)^{n/2}
    \end{split}
\end{equation}
Therefore, as we define
\begin{equation}
    s_{\ k}^{(D)}\equiv\sum_{n=0}^{k}\frac{g_{n}}{\left(k-n\right)!n!}\left(\frac{2}{2-D}\right)^{k-n}\left(\frac{4D-6}{D-2}\right)^{n/2}
\end{equation}
for $k\geq 0$, we get the following form of the action:
\begin{equation}
     S=\int\diff^{D}x\sqrt{g}e^{-\phi}\biggl[R+2\frac{D-1}{D-2}\Delta \phi+\left(\nabla \phi\right)^2-\sum_{n=0}^{+\infty}s_{\ n}^{(D)}\phi^{n}\biggr]\ .
\end{equation}
Hence, we precisely have a functional as the one described in \eqref{funk}, with:
\begin{equation}
    \alpha=1,\quad\gamma=2\frac{D-1}{D-2},\quad\beta_{n}=-s_{\ n}^{(D)}\ .
\end{equation}
The associated flow equations are:
\begin{equation}\label{acteq}
    \begin{split}
        \frac{\de g_{\mu\nu}}{\de\lambda}=&-2R_{\mu\nu}-g_{\mu\nu}\sum_{n=1}^{+\infty}ns_{\ n}^{(D)}\phi^{n-1}+4\frac{D-1}{D-2}\nabla_{\mu}\phi\nabla_{\nu}\phi-2\frac{5D-6}{D-2}\nabla_{\nu}\nabla_{\mu}\phi+\\
        &-2\frac{D-1}{D-2}g_{\mu\nu}\nabla^{2}\phi+2\frac{D-1}{D-2}g_{\mu\nu}\left(\nabla \phi\right)^2+\mathcal{L}_{\xi}g_{\mu\nu}\ ,\\
        \frac{\de \phi}{\de\lambda}=&-R-\frac{D}{2}\sum_{n=1}^{+\infty}ns_{\ n}^{(D)} \phi^{n-1}+\frac{\left(D-1\right)\left(D+2\right)}{D-2}\left(\nabla \phi\right)^{2}+\\
        &-\frac{D^{2}+4D-6}{D-2}\nabla^{2} \phi+\mathcal{L}_{\xi} \phi\ .
    \end{split}
\end{equation}
Therefore, we have derived a general expression for the geometric flow equations induced by the $D$-dimensional action for a metric-scalar system with arbitrary polynomial self-interaction terms, by defining the entropy functional as its string-frame version. 
\newpage
\section{Bubble Toy Model}\label{bubbs}
In the following section, we will start to investigate the flow behaviour of scalar bubbles under geometric flow equations. In order to do so, we will consider a simplified toy model of the bubble solutions which are expected to solve Einstein field equations in presence of a scalar field. In particular, we will relax the equations of motion in the non-trivial portions of the bubbles, namely where they interpolate among different vacua. Thus, we will allow our scalar profiles and metrics not to solve them in that limited space-time shells, in order to achieve simpler expressions. In the \textit{thin-wall} approximation, this should only account for small deviations. Hence, we will take such settings as initial conditions and make them evolve under \textit{Perleman's combined flow}
    \begin{equation}\label{form2}
    \begin{cases}
     \diff_{\lambda}g_{\mu\nu}=-2R_{\mu\nu}\ ,\\
     \diff_{\lambda}\phi= -R-\Delta\phi+\left(\nabla \phi\right)^2\ ,
    \end{cases}
\end{equation}
given by $\Bar{\mu}=\left(0,1,0,0\dots\right)$. In the next sections, after having developed an intuitive picture of the expected evolution of bubbles under geometric flows, we will move to less trivial constructions. Nevertheless, our present discussion can be valuable in outlining the main features of the investigated class of phenomena.
\subsection{Bubble construction}
Before solving Perelman's combined flow equations, we must precisely describe our bubble toy model. In order to do so, we neglect the backreaction of the scalar bubble on space-time curvature and consider a background metric structure. Then, we provide the reader with a detailed description of the scalar profile of our interest.
\subsubsection{Background Metric}
Concerning the space-time metric, we consider a static, uncharged black hole embedded in a cosmological constant background:
\begin{equation}\label{genmet}
\begin{split}
    \ds^2=&-\left(1-\frac{2M}{r}-\frac{\Lambda r^{2}}{3}\right)\diff t^2+\left(1-\frac{2M}{r}-\frac{\Lambda r^{2}}{3}\right)^{-1}\diff r^2+\\
    &+r^2\diff\theta^2+r^2\sin{\theta}^2\diff\varphi^2\ .
\end{split}
\end{equation}
Hence, we actually consider a family of metrics parametrised by the couple $\left(M,\Lambda\right)\in\R^{2}/\Z_{2}$. Having to deal with an Einstein manifold, the curvature tensor and scalar take the following, simple, forms:
\begin{equation}\label{curvspe}
    R_{\mu\nu}=\Lambda g_{\mu\nu}\ ,\quad R=4\Lambda\ .
\end{equation}
It must be stressed that, as a system is forced to evolve according to some flow equations, the metric might not be described anymore by the clear expression \eqref{genmet}. Namely, when studying a specific setting, we will have to understand whether the metric flow produces a general path in the \textit{Riemannian manifold of all Riemannian metrics} \cite{1992math......1259G} or it is simply constrained in its $\left(M,\Lambda\right)$ sub-manifold. When the second scenario is realised, the curvature tensor and scalar can be taken to have the form \eqref{curvspe} along the whole evolution in the flow parameter $\lambda$, as the flow dependence gets fully moved to $M$ and $\Lambda$.
\subsubsection{Scalar Profile}
In order to construct the scalar profile, we start by assuming it to be static and spherically symmetric. Namely, we take $\phi=\phi(r)$ in spherical coordinates $\left(t,r,\theta,\varphi\right)$. Thereafter, we consider a sphere around the origin with radius $\varrho$ and assume the scalar, far enough from its surface, to take two constant values $f_{1}$ and $f_{2}$, with $f_{1}\neq f_{2}$, inside and outside the sphere, respectively. It must be stressed that such a sphere is a $2$-dimensional space-like sub-manifold, for which the radial coordinate is fixed at $\varrho$ and time is fixed at an arbitrary value $\tau$. We can always shift $\tau$ via performing an isometry in the time direction, as both the scalar and the metric are time-independent. Concerning the scalar behaviour at the surface of the sphere, corresponding to a \textit{domain wall}, we must take it such that it interpolates properly between $f_{1}$ and $f_{2}$. By giving the domain wall a thickness equal to $2\varepsilon$, we have
\begin{equation}\label{dilaton}
    \phi(r)\equiv\begin{cases}
        \begin{split}
        f_{1}&\text{ for }r\leq\varrho-\varepsilon\\
        I(r)&\text{ for }\varrho-\varepsilon\leq r\leq\varrho+\varepsilon\\
        f_{2}&\text{ for }r\geq\varrho+\varepsilon
        \end{split}
    \end{cases}\ ,
\end{equation}
with:
\begin{equation}
    I(r)\equiv\frac{f_{2}+f_{1}}{2}+\frac{f_{2}-f_{1}}{2}\tanh{\frac{r-\varrho}{\left(\varrho+\varepsilon-r\right)\left(r-\varrho+\varepsilon\right)}}\ .
\end{equation}
At this point, both the initial metric $\Bar{g}_{\mu\nu}$ and the initial scalar $\Bar{\phi}$ have been defined appropriately. In the following discussion, we will solve Perelman's combined flow equations for some choices of the parameters $M$ and $\Lambda$, characterising $\Bar{g}_{\mu\nu}$ in specific ways.
\subsection{Minkowski background}\label{Minkowski}
We start from the simplest possible example. Namely, we take both $M=0$ and $\Lambda=0$. First of all, we observe that Minkowski space-time is characterised by a null Ricci tensor
\begin{equation}
    R_{\mu\nu}=0\ ,
\end{equation}
therefore the metric is constant along the flow. Hence, $M$ and $\Lambda$ have no evolution in the flow parameter $\lambda$. Being $R=0$, the flow equation for the scalar reduces to:
\begin{equation}\label{flsc}
    \diff_{\lambda}\phi=-\Delta\phi+\left(\nabla \phi\right)^2\ .
\end{equation}
For both $r\leq\varrho-\varepsilon$ and $r\geq\varrho+\varepsilon$, the scalar is constant in $r$. Hence, the right-hand side of \eqref{flsc} is zero: this straightforwardly implies that both $f_{1}$ and $f_{2}$ are constant along the flow.
Since we are working in a spherically symmetric setting, it can be safely expected, without loss of generality, that Perelman's flow does not spoil the spherical symmetry of $\phi$. Starting from the previous observation, we now introduce an appropriate flow ansatz, assuming that the functional form \eqref{dilaton} properly describes $\phi$ at any value of $\lambda$. Moreover, we take $\varepsilon$ to be constant along the flow. This last hypothesis is backed-up by the so-called \textit{thin-wall approximation}, which we will assume to well describe our solution for the rest of the following discussion. Namely, we impose
\begin{equation}
    \varrho\gg\varepsilon>0\ .
\end{equation}
and assume it to hold along the flow.
Hence, we can fully model the scalar flow as an evolution in the wall position $\varrho$. If bubbles will be found to shrink along the flow, the approximation of constant thickness can be expected to break down after a certain critical flow time $\lambda_{c}$. Hence, some extra care will be required. That said, we now work with in a regime in which $\varrho\gg\varepsilon$, define
\begin{equation}\label{bubb}
    \phi(r)\equiv\varphi\left(r-\varrho\right)
\end{equation}
and move the whole flow dependence to $\varrho$, with $r-\varrho\equiv x\in\left(-\varrho,+\infty\right)$.
Using the simple notation
\begin{equation}
    \Dot{\varrho}\equiv\diff_{\lambda}\varrho\ ,
\end{equation}
for the $\lambda$-derivative and writing the flow equation for the scalar profile in terms of $\varrho$ and $\varphi$, we have:
\begin{equation}
    \frac{\diff^2\varphi}{\diff x^2}=\left(\Dot{\varrho}-\frac{2}{x+\varrho}\right)\frac{\diff\varphi}{\diff x}+\left(\frac{\diff\varphi}{\diff x}\right)^2\ .
\end{equation}
By multiplying both sides by the $x$-derivative of $\varphi$, we obtain:
\begin{equation}\label{wall1flow}
\frac{1}{2}\frac{\diff}{\diff x}\left(\frac{\diff\varphi}{\diff x}\right)^2=\left(\Dot{\varrho}-\frac{2}{x+\varrho}\right)\left(\frac{\diff\varphi}{\diff x}\right)^2+\left(\frac{\diff\varphi}{\diff x}\right)^3\ .
\end{equation}
In particular, we have
\begin{equation}\label{derphi}
    \frac{\diff\varphi}{\diff x}=\begin{cases}
        0\ \text{ for }-\varrho\leq x\leq-\varepsilon\\
        K(x)\text{ for }-\varepsilon\leq x\leq\varepsilon\\
        0\ \text{ for }x\geq\varepsilon
    \end{cases}\ ,
\end{equation}
where:
\begin{equation}
    K(x)=\frac{f_{2}-f_{1}}{2}\frac{\left(x^{2}+\varepsilon^2\right)}{(x-\varepsilon)^2(x+\varepsilon)^2}\cosh^{-2}{\frac{x}{\varepsilon^2-x^2}}\equiv\frac{f_{2}-f_{1}}{2}G(x)\ .
\end{equation}
Now, we integrate \eqref{wall1flow} from $0$ to $+\infty$, in order to read-off $\Dot{\varrho}$. Thus, we are left with the following equation:
\begin{equation}\label{rhoeq}
    \Dot{\varrho}\int_{-\varepsilon}^{\varepsilon}G(x)^{2}\diff x=2\int_{-\varepsilon}^{\varepsilon}\frac{G(x)^{2}}{x+\varrho}\diff x+\frac{f_{1}-f_{2}}{2}\int_{-\varepsilon}^{\varepsilon}G(x)^{3}\diff x\ .
\end{equation}
Evidently, \eqref{rhoeq} can't be solved analytically. Anyway, we can easily observe that $G(x)$ is always positive in $(-\varepsilon,\varepsilon)$. Hence, the flow is characterised by three positive functions:
\begin{equation}\label{functions}
    \begin{split}
        &A\left(\varepsilon\right)\equiv\int_{-\varepsilon}^{\varepsilon}G(x)^{2}\diff x\ ,\\
        &B\left(\varepsilon,\varrho\right)\equiv\int_{-\varepsilon}^{\varepsilon}\frac{G(x)^{2}}{x+\varrho}\diff x\ ,\\
        &C\left(\varepsilon\right)\equiv\int_{-\varepsilon}^{\varepsilon}G(x)^{3}\diff x\ .
    \end{split}
\end{equation}
Therefore, we obtain:
\begin{equation}\label{rhof}
    \Dot{\varrho}=2\frac{B\left(\varepsilon,\varrho\right)}{A\left(\varepsilon\right)}+\frac{f_{1}-f_{2}}{2}\frac{C\left(\varepsilon\right)}{A\left(\varepsilon\right)}\ .
\end{equation}
Since we work in the $\varepsilon\ll\varrho$ regime, we have:
\begin{equation}
    B\left(\varepsilon,\varrho\right)\approx\int_{-\varepsilon}^{\varepsilon}\frac{G(x)^{2}}{\varrho}\diff x=\frac{A\left(\varepsilon\right)}{\varrho}\ .
\end{equation}
Therefore, \eqref{rhof} reduces to
\begin{equation}\label{equaa}
        \Dot{\varrho}=\frac{2}{\varrho}+\frac{f_{1}-f_{2}}{2}Q\left(\varepsilon\right)\ ,
\end{equation}
with:
\begin{equation}
   Q\left(\varepsilon\right)\equiv\frac{C\left(\varepsilon\right)}{A\left(\varepsilon\right)}\ .
\end{equation}
\begin{figure}[H]
    \centering
    \includegraphics[width=0.8\linewidth]{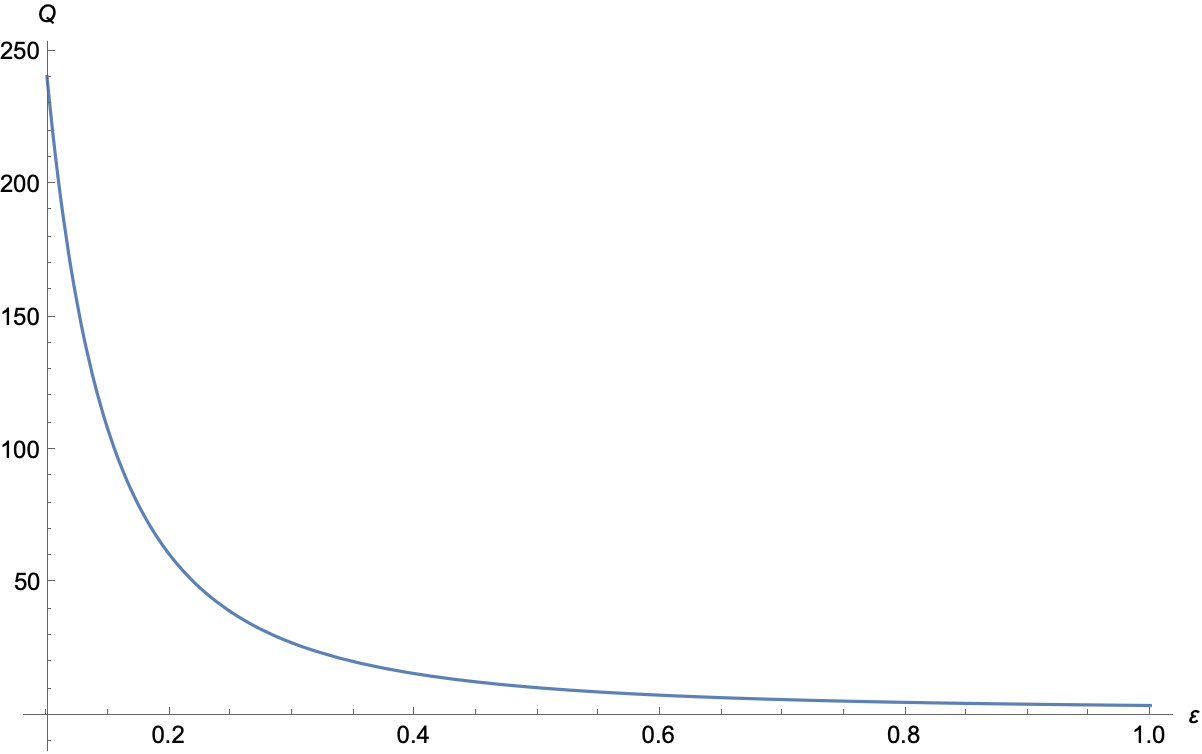}
    \caption{Plot of $Q$ for different values of the wall thickness.}
\end{figure}
\subsubsection{Bubble with a smaller dilaton value}
Assuming $f_{2}>f_{1}$, starting from a big enough value of $\varrho_{0}$ (required for our constant wall thickness approximation) and defining
\begin{equation}\label{chii}
    \chi\equiv\frac{f_{1}-f_{2}}{2}Q\left(\varepsilon\right)\ ,
\end{equation}
we can solve the flow equation for $\varrho$ as
\begin{equation}\label{rhosol}
    \varrho\left(\lambda\right)=-\frac{2}{\chi}\left\{1+W\left[-e^{-1-2\lambda\chi^2-\chi\varrho_{0}/2}\left(1+\frac{\chi\varrho_{0}}{2}\right)\right]\right\}
\end{equation}
where $W$ is Lambert's function positive branch. Strictly speaking, $W$ only admits arguments in $\left(-e^{-1},\infty\right)$. Hence, we assume to work with:
\begin{equation}\label{initial}
    \varrho_{0}\geq\frac{1}{f_{2}-f_{1}}\frac{4}{Q\left(\varepsilon\right)}\ .
\end{equation}
\begin{figure}[H]
    \centering
    \includegraphics[width=\linewidth]{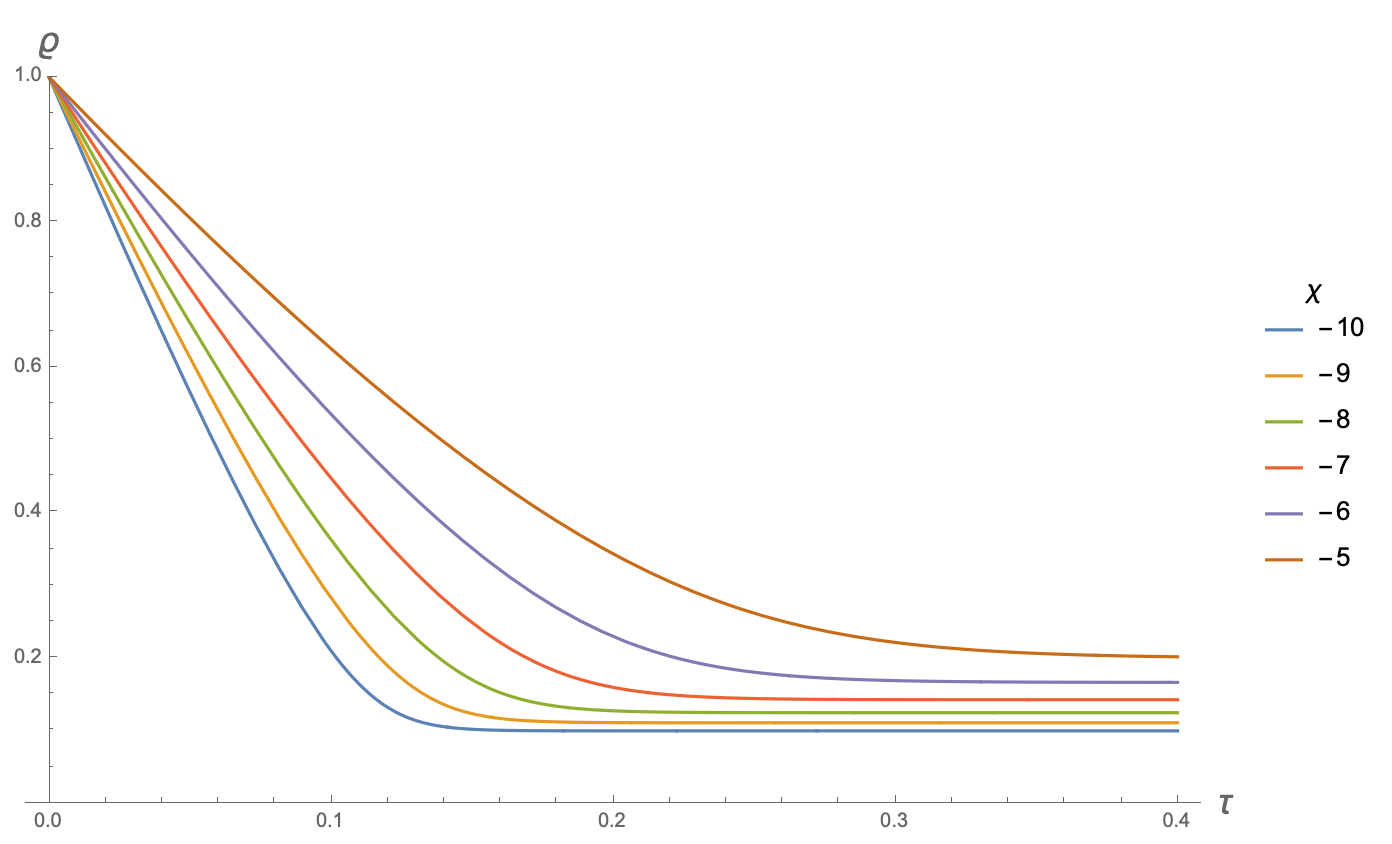}
    \caption{Flow behaviour of $\varrho$ for different values of $\chi$, with $\varrho_{0}=1$.}
\end{figure}
The regime in which the value of the dilaton inside the bubble is smaller than the one outside the bubble, namely when $f_{2}>f_{1}$, produces shrinking bubbles. In particular, starting from an initial value $\varrho_{0}$ satisfying $\eqref{initial}$, the bubbles shrink approaching the asymptotic value
\begin{equation}
    \varrho_{\infty}\equiv\frac{1}{f_{2}-f_{1}}\frac{4}{Q\left(\varepsilon\right)}\ ,
\end{equation}
for which $\Dot{\varrho}=0$ and the above bound is saturated.
\subsubsection{Bubble with a larger dilaton value}
Assuming $f_{2}<f_{1}$, starting from a big enough value of $\varrho_{0}$ (required for our constant wall thickness approximation) and defining $\chi$ as in \eqref{chii},
we can't solve for $\varrho$ as in \eqref{rhosol} since we would deal with negative arguments in Lambert's function. Nevertheless, we can straightforwardly observe that right-hand side of \eqref{equaa} is always positive. Hence, $\varrho$ is always forced to grow, asymptotically approaching a linear behaviour
\begin{equation}\label{asy}
    \Tilde{\varrho}\left(\lambda\right)\approx\chi\lambda+\varrho_{0}\ ,
\end{equation}
for which the $2/\varrho$ in \eqref{equaa} can be neglected. Therefore, when the dilaton value inside the bubble is larger than the one outside, the bubble grows indefinitely towards $\infty$. For large enough values of $\lambda$, \eqref{asy} nicely approximates the flow behaviour of its radius.
\subsection{dS/AdS background}
In the following discussion, we take $M=0$ and $\Lambda\neq 0$. 
Considering the expression \eqref{curvspe} for the Ricci tensor associate with an Einstein manifold, we have that the flow equation for the metric reduces to:
\begin{equation}\label{cosmo}
    \Dot{\Lambda}=2\Lambda^2\ .
\end{equation}
The above equation can be straightforwardly solved by: 
\begin{equation}
    \Lambda\left(\lambda\right)=\frac{\Lambda_{0}}{1-2\lambda\Lambda_{0}}\ .
\end{equation}
Therefore, \textit{de Sitter} space-time $(\Lambda_{0}>0)$ reaches a singularity $\Lambda=\infty$ at a finite flow time:
\begin{equation}\label{singul}
    \lambda_{\infty}=\frac{1}{2\Lambda_{0}}\ .
\end{equation}
\textit{Anti de Sitter} space-time $(\Lambda_{0}<0)$, on the other hand, asymptotically approaches $\Lambda=0$ as $\lambda\rightarrow\infty$. Since the sign of $\Lambda_{0}$ dramatically affects the qualitative flow behaviour, it is natural to consider the two settings separately when dealing with evolution of the scalar profile.
\subsubsection{Anti de Sitter background}
As observed above, we have that the background metric asymptotically approaches Minkowski as $\lambda\rightarrow\infty$. We now study how this evolution affects the flow of the scalar bubble at early flow times, since we can expect its $\lambda$-behaviour to gradually approach the one studied in section \eqref{Minkowski} as $\Lambda$ goes to zero. Furthermore, we have to keep in mind that early flow time analysis is more accurate since we start from a thin bubble wall regime and assume it to hold, while it might get spoiled as we move towards large $\lambda$ values. Concerning the flow behaviour of $\phi$ outside the bubble, we get
\begin{equation}\label{bbf}
    \Dot{f}_{i}=-4\Lambda=\frac{\Lambda_{0}}{1-2\lambda\Lambda_{0}}
\end{equation}
with $i=1,2$. Therefore, the $f_{i}$ values are not constant along the flow in $\lambda$, as they were in section \eqref{Minkowski}. In particular, we have:
\begin{equation}
    f_{i}\left(\lambda\right)=f_{i}\left(0\right)+2\log{\left(1-2\Lambda_{0}\lambda\right)}\ .
\end{equation}
The logarithmic behaviour is precisely produced by the fact that we asymptotically approach Minkowski space-time, for which the source term on the right-hand side of \eqref{bbf} progressively weakens. It can be observed that both $f_{1}$ and $f_{2}$ grow towards infinity. Fortunately, as we will see below, the flow equation for the bubble size will only depend on their difference, so the $\lambda$-dependent part will be factored out. Plugging \eqref{bubb}, together with the metric, into the flow equation \eqref{form2} for $\phi$ and moving the $\lambda$-dependence to $\varrho$, we get:
\begin{equation}
    \frac{\diff^2\varphi}{\diff x^2}=-4\Lambda+\left(\Dot{\varrho}-\frac{2}{x+\varrho}\right)\frac{\diff\varphi}{\diff x}+\left[1-\frac{\Lambda}{3}\left(x+\varrho\right)^2\right]\left(\frac{\diff\varphi}{\diff x}\right)^2\ .
\end{equation}
By multiplying both sides by the $x$-derivative of $\varphi$, we obtain:
\begin{equation}\label{walladsflow}
\begin{split}
\frac{1}{2}\frac{\diff}{\diff x}\left(\frac{\diff\varphi}{\diff x}\right)^2=&-4\Lambda\frac{\diff\varphi}{\diff x}+\left(\Dot{\varrho}-\frac{2}{x+\varrho}\right)\left(\frac{\diff\varphi}{\diff x}\right)^2+\\
&+\left[1-\frac{\Lambda}{3}\left(x+\varrho\right)^2\right]\left(\frac{\diff\varphi}{\diff x}\right)^3\ .
\end{split}
\end{equation}
The $x$-derivative of $\varphi$ takes the form illustrated in \eqref{derphi}. By integrating the above equation and getting rid of boundary terms, we get:
\begin{equation}
\begin{split}
    \Dot{\varrho}\int_{-\varepsilon}^{\varepsilon}G(x)^{2}\diff x=&2\int_{-\varepsilon}^{\varepsilon}\frac{G(x)^{2}}{x+\varrho}\diff x+\frac{f_{1}-f_{2}}{2}\int_{-\varepsilon}^{\varepsilon}G(x)^{3}\diff x+\\
    &-\frac{\Lambda}{3}\frac{f_{1}-f_{2}}{2}\int_{-\varepsilon}^{\varepsilon}G(x)^{3}\left(x+\varrho\right)^2\diff x\ .
\end{split}
\end{equation}
The flow in characterised by the three positive functions defined in \eqref{functions}, together with a further positive function:
\begin{equation}
    D\left(\varepsilon,\varrho\right)\equiv\int_{-\varepsilon}^{\varepsilon}G(x)^{3}\left(x+\varrho\right)^2\diff x\ .
\end{equation}
Hence, we have:
\begin{equation}\label{rhof2}
    \Dot{\varrho}=2\frac{B\left(\varepsilon,\varrho\right)}{A\left(\varepsilon\right)}+\frac{f_{1}-f_{2}}{2}\frac{C\left(\varepsilon\right)}{A\left(\varepsilon\right)}-\frac{\Lambda}{3}\frac{f_{1}-f_{2}}{2}\frac{D\left(\varepsilon,\varrho\right)}{A\left(\varepsilon\right)}\ .
\end{equation}
Given the thin-wall approximation, we get
\begin{equation}
    D\left(\varepsilon,\varrho\right)\approx\int_{-\varepsilon}^{\varepsilon}G(x)^{3}\varrho^2\diff x=C\left(\varepsilon\right)\varrho^2
\end{equation}
and
\begin{equation}\label{eqbb}
    \Dot{\varrho}=\frac{2}{\varrho}+\frac{f_{1,0}-f_{2,0}}{2}\left(1-\frac{\Lambda\varrho^2}{3}\right)Q\left(\varepsilon\right)\ ,
\end{equation}
with:
\begin{equation}
   Q\left(\varepsilon\right)\equiv\frac{C\left(\varepsilon\right)}{A\left(\varepsilon\right)}\ .
\end{equation}
Introducing, once more, the parameter $\chi$, we obtain:
\begin{equation}\label{eqeq}
       \Dot{\varrho}=\frac{2}{\varrho}+\chi\left(1-\frac{\Lambda\varrho^2}{3}\right)\ .
\end{equation}
We observe that, for Anti de Sitter space-time, $\Lambda$ does nothing more than enhancing the contribution of $\chi$ to the source term on the right-hand side of \eqref{eqeq}, particularly for early flow times and large values of $\varrho$. Therefore, there is no qualitative difference with the analysis developed in section \eqref{Minkowski}. Since $\Lambda\rightarrow 0$, even the asymptotic radius of shrinking bubbles in the $\chi<0$ regime is unchanged from the discussion developed for Minkowski background.
\subsubsection{de Sitter background}
Since its formal derivation coincides with the one developed for Anti de Sitter space-time, we find ourselves with the same thin-wall regime flow equation:
\begin{equation}
       \Dot{\varrho}=\frac{2}{\varrho}+\chi\left(1-\frac{\Lambda\varrho^2}{3}\right)\ .
\end{equation}
As discussed before, the evolution in $\lambda$ reaches a finite-time singularity at
\begin{equation}
    \lambda_{\infty}=\frac{1}{2\Lambda_{0}}\ ,
\end{equation}
where $\Lambda\rightarrow\infty$ and $f_{i}\rightarrow-\infty$, for $i=1,2$. $\chi$ is still constant along the flow. Nevertheless, it can be easily observed that this setting is way richer and more subtle that the ones studied before. Therefore, different regions in the space of parameters $\left(\varrho_{0},\chi,\Lambda_{0}\right)$ produce a wide variety of flow behaviours and must be analysed separately. Depending on the values of the parameters, the bubbles can steadily grow, shrink or follow non-monotonic behaviours. 
Taking $\Lambda_{0}=1/2$, so that $\lambda_{\infty}=1$, we study the evolution for different values of $\chi$ and $\varrho_{0}$.
\begin{figure}[H]
    \centering
    \includegraphics[width=\linewidth]{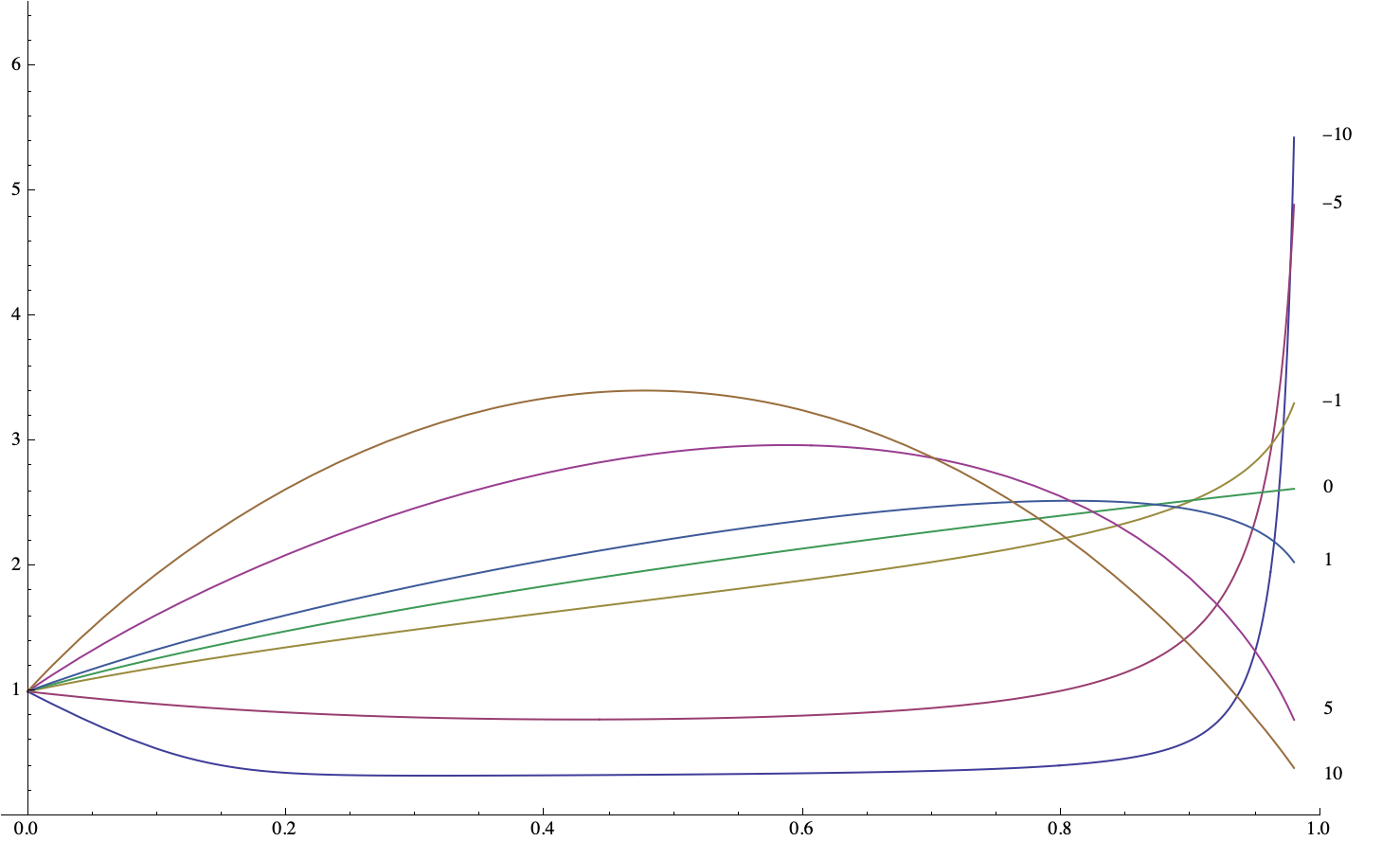}
    \caption{$\varrho\left(\lambda\right)$ with $\varrho_{0}=1$, at different values of $\chi$.}
    \includegraphics[width=\linewidth]{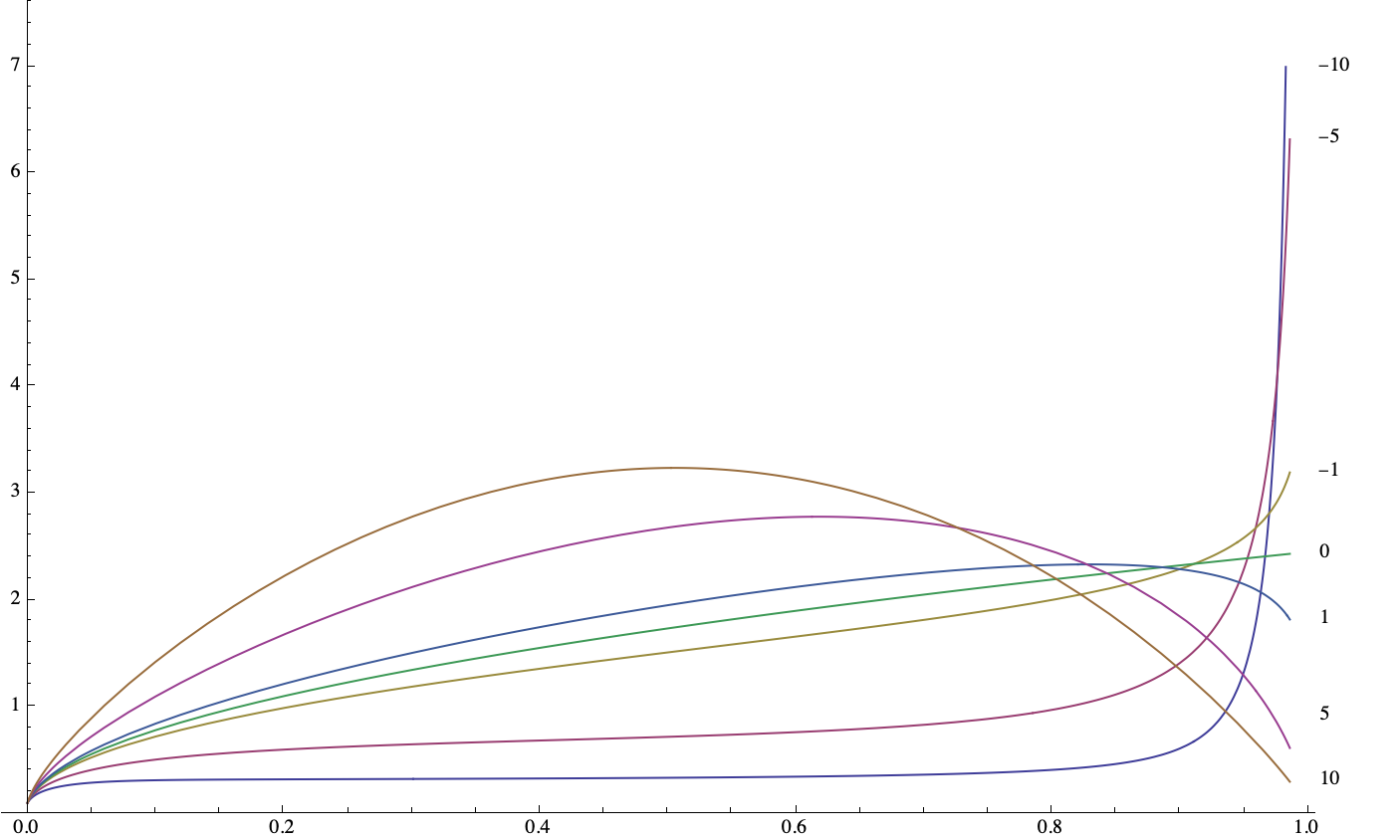}
    \caption{$\varrho\left(\lambda\right)$ with $\varrho_{0}=0.1$, at different values of $\chi$.}
    \end{figure}
\begin{figure}[H]
    \centering
    \includegraphics[width=\linewidth]{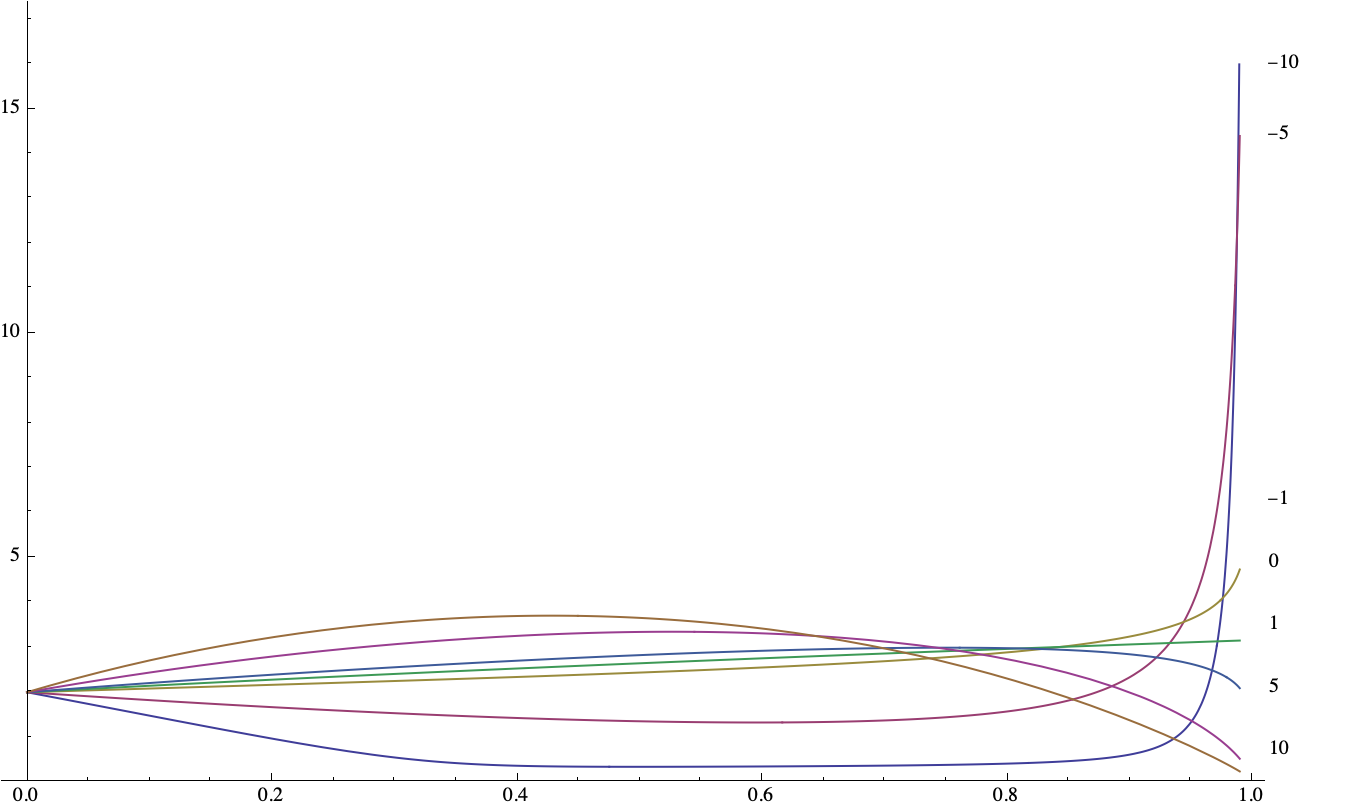}
    \caption{$\varrho\left(\lambda\right)$ with $\varrho_{0}=2$, at different values of $\chi$.}
\end{figure}
Concerning the above examples, they clearly highlight how many different flow behaviours can be produced by appropriately choosing the initial values of the solution parameters. In particular, there are regions of the parameters space for which the bubble radius exceeds that of de Sitter space-time along the flow. This clearly generates an inconsistency, as the bubble would pass through the cosmological horizon. This feature might signal the fact that the only allowed choices for the initial parameters are the ones that do not lead us into such a situation.
\subsection{Schwarzschild Background}
In the following discussion, we work with Schwarzschild metric in standard coordinates"
\begin{equation}\label{schw}
    \ds^2=-\left(1-\frac{2M}{r}\right)\diff t^2+\left(1-\frac{2M}{r}\right)^{-1}\diff r^2+r^2\diff\theta^2+r^2\sin{\theta}^2\diff\phi^2\ .
\end{equation}
Namely, we take $\Lambda=0$ and $M>0$. Therefore, we will only consider bubbles larger than the black hole horizon radius $r_{h}=2M$. Concerning the scalar profile, we stick to the form presented in \eqref{dilaton} and take the ansatz \eqref{bubb}. Furthermore, we assume the thin wall approximation $\varrho\gg\varepsilon$. For a radial function, the Laplacian is:
\begin{equation}
    \Delta\phi=\frac{1}{\sqrt{-g}}\de_{\mu}\left(\sqrt{-g}g^{\mu\nu}\de_{\nu}\phi\right)=\left(1-\frac{2M}{r}\right)\frac{\diff^2\phi}{\diff r^2}+2\frac{r-M}{r^2}\frac{\diff\phi}{\diff r}\ .
\end{equation}
Hence, the flow equation \eqref{form2} for $f$ takes the form:
\begin{equation}\label{equasc}
    \Dot{\phi}=-\left(1-\frac{2M}{r}\right)\frac{\diff^2\phi}{\diff r^2}-2\frac{r-M}{r^2}\frac{\diff\phi}{\diff r}+\left(1-\frac{2M}{r}\right)\left(\frac{\diff\phi}{\diff r}\right)^2\ .
\end{equation}
By plugging \eqref{bubb} in \eqref{equasc}, we have:
\begin{equation}
     \frac{\diff^2\varphi}{\diff x^2}=\left(1-\frac{2M}{x+\varrho}\right)^{-1}\left(\Dot{\varrho}-2\frac{x+\varrho-M}{\left(x+\varrho\right)^2}\right)\frac{\diff\varphi}{\diff x}+\left(\frac{\diff\varphi}{\diff x}\right)^2\ .
\end{equation}
Since $x\in\left(-\varepsilon,\varepsilon\right)$, we have $x+\varrho\approx\varrho$. Moreover, we work with $\varrho>2M$. Hence, the flow equation can be approximated as:
\begin{equation}
    \frac{\diff^2\varphi}{\diff x^2}=\frac{\varrho}{\varrho-2M}\left(\Dot{\varrho}+2\frac{M-\varrho}{\varrho^2}\right)\frac{\diff\varphi}{\diff x}+\left(\frac{\diff\varphi}{\diff x}\right)^2\ .
\end{equation}
By multiplying both sides by $\diff\varphi/\diff x$ and integrating as was done in section \ref{Minkowski}, we have:
\begin{equation}
   \Dot{\varrho}\int_{-\varepsilon}^{\varepsilon}\left(\frac{\diff\varphi}{\diff x}\right)^2\diff x=2\frac{\varrho-M}{\varrho^2}\int_{-\varepsilon}^{\varepsilon}\left(\frac{\diff\varphi}{\diff x}\right)^2\diff x-\frac{\varrho-2M}{\varrho}\int_{-\varepsilon}^{\varepsilon}\left(\frac{\diff\varphi}{\diff x}\right)^3\diff x\ .
\end{equation}
With the notation introduced in section $\ref{Minkowski}$, we obtain:
\begin{equation}
    \Dot{\varrho}=2\frac{\varrho-M}{\varrho^2}+\chi\frac{\varrho-2M}{\varrho}\ .
\end{equation}
It can be observed that, when $M=0$, the above reduces to \eqref{equaa}. Moreover, by studying the near-horizon behaviour of the flow, it can be shown that the $\varrho>2M$ assumption is conserved along the flow. Namely, by taking $\varrho=2M+\mu$, with $\mu\ll 2M$, we have:
\begin{equation}
    \Dot{\varrho}\approx\frac{1}{2M}>0\ .
\end{equation}
Hence, starting from $\varrho_{0}>2M$ we are forced to stay in the $\varrho>2M$ regime. Namely, if taken outside the black hole, the wall can't cross the horizon along the flow and stays where the metric \eqref{schw} is properly defined. Hence, the presence of a non-zero $M$ cannot produce the problematic behaviour encountered when analysing de Sitter background. Concerning fixed points of the flow, we observe:
\begin{equation}
    \Dot{\varrho}=0\quad\Longrightarrow\quad 2\frac{\varrho-M}{\varrho^2}+\chi\frac{\varrho-2M}{\varrho}=0
\end{equation}
Thus, it seems like we have fixed points $\Bar{\varrho}_{\pm}$ of the flow solving
\begin{equation}
    \Bar{\varrho}^2+2\frac{1-M\chi}{\chi}\Bar{\varrho}-\frac{2M}{\chi}=0\ ,
\end{equation}
but we still have to discuss whether any of the two lies outside the black hole event horizon, which is the region we are interested in. We have:
\begin{equation}
    \Bar{\varrho}_{\pm}=M-\frac{1}{\chi}\pm\frac{\sqrt{M^2\chi^2+1}}{\chi}\ .
\end{equation}
We observe that, for $M>0$, only $\Bar{\varrho}_{-}$ can be bigger than $2M$. Moreover, we can only achieve this with $\chi<0$. Therefore:
\begin{itemize}
    \item $\chi>0$: There's no fixed point of the flow.
    \item $\chi<0$: The bubble radius $\Bar{\varrho}_{-}$ is a fixed point of the flow.
\end{itemize}
\begin{figure}[H]
    \centering
    \includegraphics[width=0.8\linewidth]{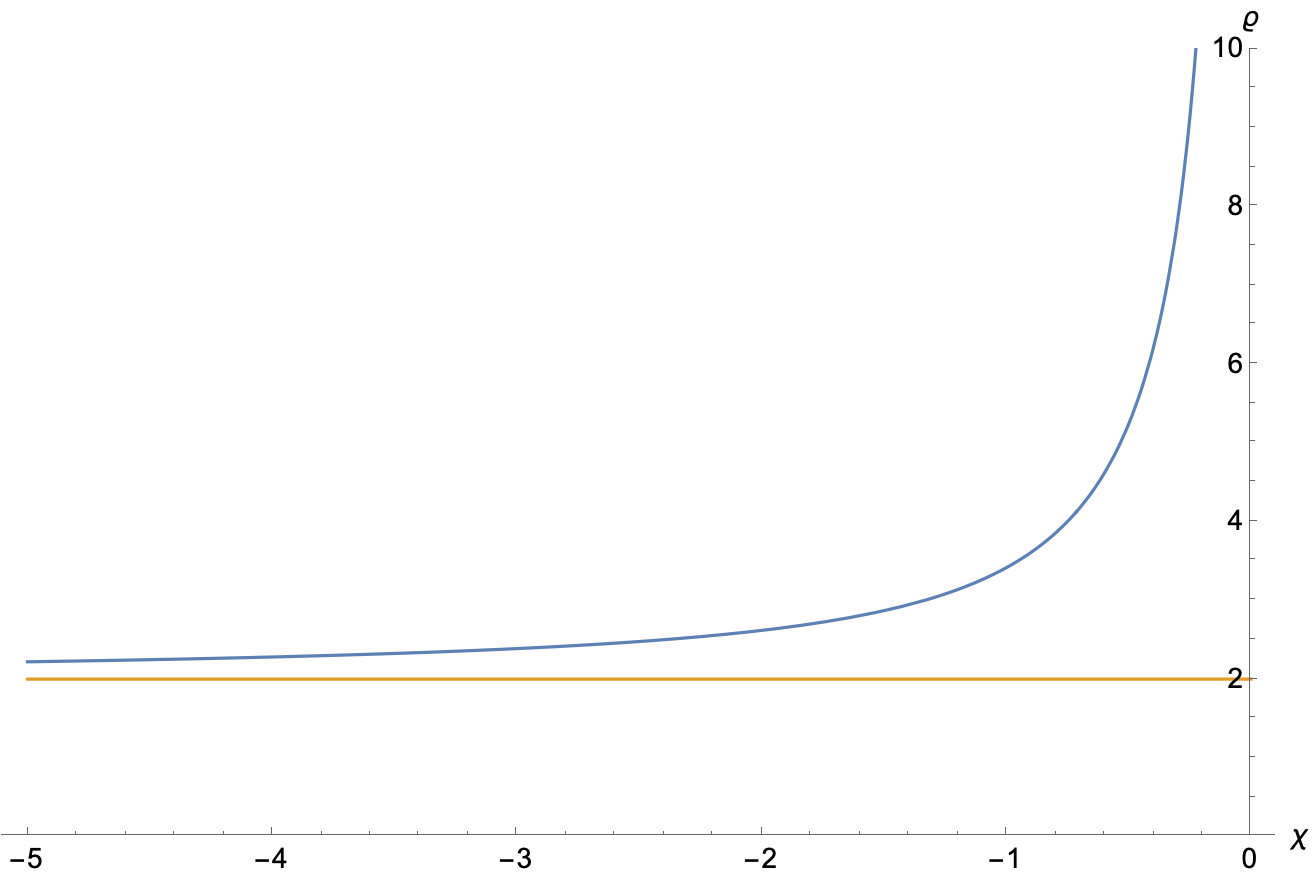}
    \caption{Plot of $\Bar{\varrho}_{-}$ (blue line), for $M=1$, against negative values of $\chi$. As can be clearly observed, it is bigger than the horizon radius (orange line).}
\end{figure}
We now study the $\chi<0$ case in detail. As discussed above, we have a fixed point at:
\begin{equation}
       \Bar{\varrho}_{-}=M-\frac{1}{\chi}-\frac{\sqrt{M^2\chi^2+1}}{\chi}\ .
\end{equation}
It can be easily observed that $\Dot{\varrho}>0$ for $2M<\varrho<\Bar{\varrho}_{-}$ and $\Dot{\varrho}<0$ for $\Bar{\varrho}_{-}<\varrho$. Namely, we always tend towards the fixed point at $\Bar{\varrho}_{-}$.\\
For $\chi>0$, instead, we simply have that $\Dot{\varrho}>0$ for every value of $\varrho$. Therefore, the bubble is forced to grow indefinitely.
\section{Normalised Flow of Einstein Bubbles}\label{norms}
In the following section, we consider space-time kinematical states with cosmological constant bubbles. Thus, we focus on Anti de Sitter space-time with a metric bubble as an initial condition, switch on the \textit{normalised} evolution equations in $\lambda$ and compute the distance along the flow with the appropriate entropy functional. The whole discussion will be performed in Euclidean signature. 
\subsection{Entropy Functional}
Let $\mathcal{M}$ be a $D$-dimensional Euclidean manifold, on which a Riemannian metric $g_{\mu\nu}$ and a scalar field $\varphi$ are defined. Let's consider the entropy functional
\begin{equation}\label{FNC}
    \mathcal{F}\left[g,\varphi\right]=\int\diff^{D}x\sqrt{g}e^{-\varphi}\frac{R}{g^{1/D}}\ ,
\end{equation}
on which we perform volume-preserving variations of the fields, such that:
\begin{equation}
    \delta\left(\sqrt{g}e^{-\varphi}\right)=0\ .
\end{equation}
Namely, we impose:
\begin{equation}
    \delta\varphi=\frac{1}{2}g_{\mu\nu}\delta g^{\mu\nu}\ .
\end{equation}
From the entropy functional \eqref{FNC}, we can straightforwardly obtain the flow equations:
\begin{equation}
    \begin{split}\label{FEE}
        \frac{\de g_{\mu\nu}}{\de\lambda}&=\frac{R}{D}g_{\mu\nu}-R_{\mu\nu}\ ,\\
        \frac{\de\varphi}{\de\lambda}&=0\ .
    \end{split}
\end{equation}
Therefore, $\varphi$ does not enter the equation for $g_{\mu\nu}$ and is frozen along the flow. Given the fact that it fully decouples, it will be neglected in the following discussion. Any Einstein manifold, for which $R_{\mu\nu}D=Rg_{\mu\nu}$, is a fixed point of the above flow equation. Now, we can move on to the explicit construction of a cosmological constant bubble in Anti de Sitter space-time.

\subsection{AdS Bubble in AdS}
Let $\mathcal{M}$ be a $D$-dimensional non-compact manifold, endowed with coordinates $\left(t,r,x^i\right)$ and with a \textit{Euclidean} metric tensor $g_{\mu\nu}$. In particular, let's consider two values $r_0$ and $r_1$ of the radial coordinate, with $r_0<r_1$, such that:
\begin{enumerate}
    \item For $0<r<r_0$, the metric is that of Euclidean Anti de Sitter space-time with cosmological constant $\Lambda_0$;
    \item For $r_0<r<r_1$, the metric is that of a domain wall interpolating among two values $\Lambda_0$ and $\Lambda_1$ of the cosmological constant;
    \item For $r>r_1$, the metric is that of Euclidean Anti de Sitter space-time with cosmological constant $\Lambda_1$.
\end{enumerate}
We introduce two positive parameters
\begin{equation}
    2\varepsilon\equiv r_1-r_0\ ,\quad \varrho\equiv\frac{r_1+r_0}{2}\ ,
\end{equation}
respectively referring to the size and the centre of the domain wall.
Furthermore, we set $D=4$ for the sake of simplicity. Hence, we work with a metric tensor of the form
\begin{equation}\label{ADS}
    \ds^{2}=\frac{r^2}{\alpha^2\left(r\right)}\diff t^2+\frac{\alpha^2\left(r\right)}{r^2}\diff r^2+\frac{r^2}{\alpha^2\left(r\right)}\diff\Bar{x}^2\ ,
\end{equation}
where
\begin{equation}
    \alpha(r)\equiv\begin{cases}
        \begin{split}
        \alpha_{0}&\text{ for }0\leq r\leq r_0\\
        G(r)&\text{ for }r_0\leq r\leq r_1\\
        \alpha_{1}&\text{ for }r_1\leq r<\infty
        \end{split}
    \end{cases}\ ,
\end{equation}
and the interpolating function is:
\begin{equation}
    G(r)\equiv\frac{\alpha_1+\alpha_0}{2}+\frac{\alpha_1-\alpha_0}{2}\tanh{\frac{r-\varrho}{\left(\varepsilon+\varrho-r\right)\left(r-\varrho+\varepsilon\right)}}\ .
\end{equation}
Given that we want to move the full flow dependence to the function $\alpha(x)$, we can translate \eqref{FEE} into a single equation for the scalar curvature and simplify the problem. Indeed, we get:
\begin{equation}\label{SRF}
    \frac{\de R}{\de\lambda}=R^{\mu\nu}R_{\mu\nu}-\frac{\Delta R}{4}-\frac{R^2}{4}\ .
    \end{equation}
Considering the flow away from $\varrho$, we see that $\alpha_0$ and $\alpha_1$ are fixed in $\lambda$. Hence, the full flow dependence can be moved to $\varepsilon$ and $\varrho$. At this point, we move to the \textit{thin-wall} approximation, assume the whole flow dependence can be pushed to $\varrho$, introduce a new radial variable $x\equiv r-\varrho$ and take:
\begin{equation}
   \alpha\left(r\right)\equiv\psi\left(r-\varrho\right)=\psi\left(x\right)\ .
\end{equation}
Referring with $\Dot{f}$ to $\lambda$-derivatives and with $f'$ to $x$-derivatives, we have:
\begin{equation}
    \frac{\de\alpha}{\de r}=\psi'\ ,\quad\Dot{\alpha}=-\Dot{\varrho}\cdot\psi'\ .
\end{equation}
From the above expression for the metric, we get:
\begin{equation}
    R=\frac{6}{\psi^4} \left\{\psi\left(x+\varrho\right)\left[6 \psi'+\left(x+\varrho\right)\psi''\right]-2 \psi^2-4 \left(x+\varrho\right)^2 \psi'^2\right\}\ .
\end{equation}
By taking the $\lambda$-derivative, we obtain:
\begin{equation}
\begin{split}
    \Dot{R}&=\Dot{\varrho}\frac{6}{\psi^4} \biggl\{4\psi'\left(x+\varrho\right)6 \psi'+4\psi'\left(x+\varrho\right)^2\psi''-8\psi' \psi-16 \left(x+\varrho\right)^2 \frac{\psi'^3}{\psi}+\\
    &\quad-\psi'^2\left(x+\varrho\right)6 -\psi'\left(x+\varrho\right)^2\psi''+\psi6 \psi'+\psi\left(x+\varrho\right)\psi''+\\
    &\quad-6\psi''\psi\left(x+\varrho\right)-\psi\left(x+\varrho\right)^2\psi'''+\psi''\psi\left(x+\varrho\right)+4 \psi\psi'+\\
    &\quad-8 \left(x+\varrho\right) \psi'^2+8 \left(x+\varrho\right)^2 \psi'\psi''\biggr\}=\\
    &=\Dot{\varrho}\frac{6}{\psi^4} \biggl\{10\psi'\left(x+\varrho\right) \psi' +2 \psi\psi'-4\psi''\psi\left(x+\varrho\right)-\psi\left(x+\varrho\right)^2\psi'''+\\
    &\quad+11\left(x+\varrho\right)^2 \psi'\psi''-16 \left(x+\varrho\right)^2 \frac{\psi'^3}{\psi}\biggr\}=\\
    &=\Dot{\varrho}\left[C_{2}\left(x\right)\left(x+\varrho\right)^2+C_{1}\left(x\right)\left(x+\varrho\right)+C_{0}\left(x\right)\right]\ .
\end{split}
\end{equation}
In the above, we have introduced:
\begin{equation}
    \begin{split}
        C_{2}\left(x\right)&\equiv\frac{6}{\psi^5}\left[11\psi\psi'\psi''-\psi^2\psi'''-16\psi'^3\right]\ ,\\
        C_{1}\left(x\right)&\equiv\frac{12}{\psi^4}\left[5\psi'^2-2\psi\psi''\right]\ ,\\
        C_{0}\left(x\right)&\equiv\frac{12}{\psi^3}\psi'\ .
    \end{split}
\end{equation}
By collecting $\varrho$ terms in $\Dot{R}$, we are left with
\begin{equation}
    \Dot{R}=\Dot{\varrho}\left[D_{2}\left(x\right)\varrho^2+D_{1}\left(x\right)\varrho+D_0\left(x\right)\right]\ ,
\end{equation}
where:
\begin{equation}
    \begin{split}
        D_{2}\left(x\right)&\equiv C_{2}\left(x\right)\ ,\\
        D_{1}\left(x\right)&\equiv 2xC_{2}\left(x\right)+C_{1}\left(x\right)\ ,\\
        D_{0}\left(x\right)&\equiv x^2C_{2}\left(x\right)+xC_{1}\left(x\right)+C_0\left(x\right)\ .
    \end{split}
\end{equation}
Now, we want to write in a similar fashion the \textit{RHS}
\begin{equation}
    K\equiv R^{\mu\nu}R_{\mu\nu}-\frac{\Delta R}{4}-\frac{R^2}{4}
\end{equation}
of the scalar curvature flow equation. We get:
\begin{equation}
\begin{split}
    K&=-\frac{3 \left(x+\varrho\right)}{2 \psi^8} \biggl\{16 \left(x+\varrho\right)^2 \psi'^2 \psi \left[20 \psi'+9 \left(x+\varrho\right)
   \psi''\right]-\\
   &\quad+\left(x+\varrho\right) \psi^2 \bigl[13 \left(x+\varrho\right)^2 \psi''^2+2 \left(x+\varrho\right) \psi' \bigl(91
   \psi''+\\
   &\quad+9 \left(x+\varrho\right) \psi'''\bigr)+208 \psi'^2\bigr]+\psi^3 \bigl[\left(x+\varrho\right)\bigl(\left(x+\varrho\right)^2 \psi''''+\\
   &\quad+50 \psi''+14 \left(x+\varrho\right) \psi'''\bigr)+40
   \psi'\bigr]-152 \left(x+\varrho\right)^3 \psi'^4\biggr\}=\\
   &=\sum_{k=1}^{4}G_{k}\left(x\right)\left(x+\varrho\right)^k\ .
\end{split}
\end{equation}
In the above expression, we've introduced:
\begin{equation}
    \begin{split}
        G_{4}\left(x\right)&\equiv-\frac{3}{2\psi^8}\left\{144 \psi'^2 \psi  \psi''-\psi^2 13\psi''^2+\psi^2 18  \psi'   \psi'''-152  \psi'^4+ \psi^3\psi''''\right\}\ ,\\
        G_{3}\left(x\right)&\equiv-\frac{3}{2\psi^8}\left\{\psi^3 14 \psi'''+320 \psi'^3\psi+182\psi^2   \psi' 
   \psi''\right\}\ ,\\
        G_{2}\left(x\right)&\equiv-\frac{3}{2\psi^8}\left\{\psi^2 208 \psi'^2+\psi^3 50 \psi''\right\}\ ,\\
        G_{1}\left(x\right)&\equiv-\frac{60}{\psi^8}\psi^3 
   \psi'\ .
    \end{split}
\end{equation}
At this point, we collect powers of $\varrho$ and get
\begin{equation}
    K\equiv N_{4}\left(x\right)\varrho^4+N_{3}\left(x\right)\varrho^3+N_{2}\left(x\right)\varrho^2+N_{1}\left(x\right)\varrho+N_{0}\left(x\right)\ ,
\end{equation}
where we have defined the functions:
\begin{equation}
    \begin{split}
        N_{4}\left(x\right)&\equiv G_{4}\left(x\right)\ ,\\
        N_{3}\left(x\right)&\equiv 4xG_{4}\left(x\right)+G_{3}\left(x\right)\ ,\\
        N_{2}\left(x\right)&\equiv 6x^2 G_{4}\left(x\right)+3x G_{3}\left(x\right)+G_{2}\left(x\right)\ ,\\
        N_{1}\left(x\right)&\equiv 4x^3 G_{4}\left(x\right)+3x^2 G_{3}\left(x\right)+2xG_{2}\left(x\right)+G_{1}\left(x\right)\ ,\\
        N_{0}\left(x\right)&\equiv x^4 G_{4}\left(x\right)+x^3 G_{3}\left(x\right)+x^2 G_{2}\left(x\right)+x G_{1}\left(x\right)\ .
    \end{split}
\end{equation}
Observing that all $D_{i}\left(x\right)$ and $N_{i}\left(x\right)$ are zero for $x\not\in\left[-\varepsilon,+\varepsilon\right]$, we can introduce the integrated constants:
\begin{equation}
    \mathcal{D}_{i}\equiv\int_{-\varepsilon}^{+\varepsilon}D_{i}\left(x\right)\diff x\ ,\quad \mathcal{N}_{i}\equiv\int_{-\varepsilon}^{+\varepsilon}N_{i}\left(x\right)\diff x\ .
\end{equation}
Therefore, the flow equation can simply be written as:
\begin{equation}
    \Dot{\varrho}\sum_{i=0}^{2}\mathcal{D}_i\varrho^i=\sum_{j=0}^{4}\mathcal{N}_j\varrho^j\ .
\end{equation}
By defining
\begin{equation}
    \mathcal{S}\equiv\frac{\mathcal{N}_0}{\mathcal{D}_0}\ ,\quad\mathcal{L}\equiv\frac{\mathcal{N}_4}{\mathcal{D}_2}\ ,
\end{equation}
we have that the \textit{large} $\varrho$ behaviour is controlled by sign of $\mathcal{L}$, while the \textit{small} $\varrho$ behaviour is controlled by sign of $\mathcal{S}$. In order to investigate the general flow behaviour, we fix $\alpha_1\equiv 1$ without loss of generality, as it only corresponds to setting a scale, and study the sign of $\Dot{\varrho}$ as a function of $\varrho$ and $\alpha_0$. It can be straightforwardly observed that the value of $\varrho$ doesn't really affect the sign of $\Dot{\varrho}$. Therefore, given the values of $\alpha_0$ and $\alpha_1$, the bubble follows either a monotonic growing or monotonic shrinking behaviour, regardless of the initial value $\varrho_0$ of its radius. Thus, the full qualitative dynamics can be captured by a plot of the sign of $\Dot{\varrho}$ against $\alpha_0$.
    \begin{figure}[H]
    \centering
    \includegraphics[width=0.9\linewidth]{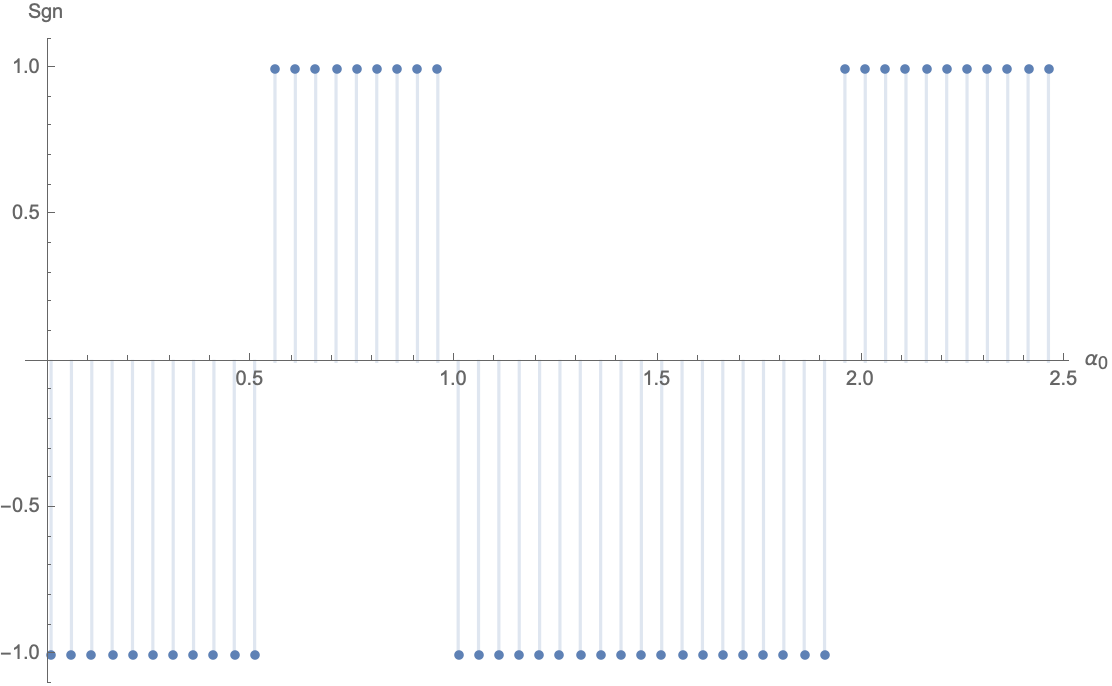}
    \caption{Plot of the sign of $\Dot{\varrho}$ against the value of $\alpha_0$, when setting $\alpha_1\equiv 1$. Regardless of the initial radius $\varrho_0$, the bubble will either grow or shrink monotonically.}
\end{figure}
\subsubsection{Distances along the Flow}
Concerning the evaluation of distances along the flow, at $D=4$, we can compute it with the formula
\begin{equation}
    \Delta\left(\varrho,\varrho_0\right)=\log{\frac{\mathcal{F}\left(\varrho\right)}{\mathcal{F}\left(\varrho_0\right)}}\ ,
\end{equation}
where $\mathcal{F}\left(\varrho\right)$ is the entropy functional computed for the metric characterised by radius $\varrho$.
Introducing an upper radial cut-off $\Omega$ and extracting as $\mathcal{F}_0$ the non-radial part of the entropy functional, which will anyway simplify in $\Delta$, we get:
\begin{equation}
    \mathcal{F}_{\Omega}=\mathcal{F}_0\int_{0}^{\Omega}\diff r\ \frac{6r}{\alpha^5}\left[\alpha r\left(6\alpha'+r\alpha''\right)-2\alpha^2-4r^2\alpha'^2\right]\equiv\mathcal{F}_0\cdot\mathcal{I}_{\Omega}\ .
\end{equation}
The above integral can be split as:
\begin{equation}
\begin{split}
    \mathcal{I}_{\Omega}&=\int_{\varrho-\varepsilon}^{\varrho+\varepsilon}\diff r\ \frac{6r}{\alpha^5}\left[\alpha r\left(6\alpha'+r\alpha''\right)-2\alpha^2-4r^2\alpha'^2\right]+\\
    &\quad -\int_{0}^{\varrho-\varepsilon}\diff r\ \frac{12r}{\alpha_0^3}-\int_{\varrho+\varepsilon}^{\Omega}\diff r\ \frac{12r}{\alpha_1^3}=\\
    &=\int_{\varrho-\varepsilon}^{\varrho+\varepsilon}\diff r\ \frac{6r}{\alpha^5}\left[\alpha r\left(6\alpha'+r\alpha''\right)-2\alpha^2-4r^2\alpha'^2\right]+\\
    &\quad -\frac{6}{\alpha_0^3}\left(\varrho-\varepsilon\right)^2-\frac{6}{\alpha_1^3}\left[\Omega^2-\left(\varrho+\varepsilon\right)^2\right]\ .
\end{split}
\end{equation}
By working in the \textit{thin-wall} approximation, we have $\varrho+\varepsilon\sim\varrho-\varepsilon\sim\varrho$. Hence, we get:
\begin{equation}
    \mathcal{I}_{\Omega}\left(\varrho\right)\sim-\frac{6}{\alpha_0^3}\varrho^2-\frac{6}{\alpha_1^3}\left[\Omega^2-\varrho^2\right]\ .
\end{equation}
Hence, we can give a rough estimate of $\Delta_{\Omega}$, where $\Omega$ must be sent to $\infty$ \textit{after} having studied the limits in $\varrho_f$, as:
\begin{equation}
    \Delta_{\Omega}\sim\log{\frac{\mathcal{I}_{\Omega}\left(\varrho\right)}{\mathcal{I}_{\Omega}\left(\varrho_0\right)}}\ .
\end{equation}
We only focus on the $\varrho_f\mapsto\infty$ limit, since the shrinking behaviour leads to a breakdown of the thin-wall approximation. Indeed, the infinite radius bubble limit sits at infinite distance even before sending $\Omega$ to infinity, with:
\begin{equation}
    \Delta_{\Omega}\propto\log{\varrho}\ .
\end{equation}
By keeping the dependence on the radial cut-off $\Omega$, we are trying to discuss the \textit{intensive} distance in the same spirit as the one motivating the quantum field theoretic version of the \textit{information metric} presented in \cite{stout2021infinite}.
\newpage
\section{On-Shell construction}\label{gfaa}
Let's consider a $D$-dimensional Riemannian manifold $\mathcal{M}$, over which a metric tensor $g_{\mu\nu}$ and a scalar $\varphi$ are defined. Furthermore, let their dynamics be governed by the Euclidean action:
\begin{equation}
    \mathcal{S}\left[g,\varphi\right]\equiv\frac{1}{2\kappa}\int\diff^Dx\sqrt{g}R+\int\diff^Dx\sqrt{g}\left(\frac{1}{2}\nabla^{\mu}\varphi\nabla_{\mu}\varphi+\beta_{n}\varphi^{n}+\beta_{m}\varphi^{m}\right)\ .
\end{equation}
The energy-momentum tensor associated to the matter part $\mathcal{S}_m\left[g,\varphi\right]$ of the action can be obtained as:
\begin{equation}
    T_{\mu\nu}\equiv\frac{1}{\sqrt{g}}\frac{\delta\mathcal{S}_m}{\delta g^{\mu\nu}}=\frac{1}{2}\nabla_{\mu}\varphi\nabla_{\nu}\varphi-\frac{1}{2}g_{\mu\nu}\left(\nabla_{\alpha}\varphi\nabla^{\alpha}\varphi+\beta_{n}\varphi^{n}+\beta_{m}\varphi^{m}\right)\ .
\end{equation}
Thus, the equations of motion associated to $\mathcal{S}\left[g,\varphi\right]$ are:
\begin{equation}
        R_{\mu\nu}-\frac{1}{2}g_{\mu\nu}R=\kappa T_{\mu\nu}\ ,\quad \nabla^2\varphi=n\beta_{n}\varphi^{n-1}+m\beta_{m}\varphi^{m-1}\ .
\end{equation}
Now, we consider the simplest possible non-trivial ansatz we can think of. Namely, we assume the geometry to be that of an \textit{Einstein manifold} and the scalar field to be \textit{constant}. Indeed, these assumptions can be phrased as:
\begin{equation}
    R_{\mu\nu}=\frac{R}{D}g_{\mu\nu}\ ,\quad \nabla_{\mu}\varphi=0\ .
\end{equation}
Plugging such conditions into the equations of motion, we are simply left with:
\begin{equation}
\begin{split}
R=&\frac{D\kappa}{D-2}\left(\beta_{n}\varphi^{n}+\beta_{m}\varphi^{m}\right)\ ,\\
    \varphi=&\left(-\frac{m\beta_m}{n\beta_n}\right)^{\frac{1}{n-m}}\ . 
\end{split}
\end{equation}
By inserting the expression for $\varphi$ into $R$, we get:
\begin{equation}
    R=\frac{D\kappa}{D-2}\left[\beta_{n}\left(-\frac{m\beta_m}{n\beta_n}\right)^{\frac{n}{n-m}}+\beta_{m}\left(-\frac{m\beta_m}{n\beta_n}\right)^{\frac{m}{n-m}}\right]\ .
\end{equation}
The \textit{most general} way in which we can understand the reduced moduli space we get when considering Einstein manifolds with a space-time constant scalar is that of a $6$-dimensional object, parametrised by $\left(D,\kappa,n,m,\beta_{m},\beta_n\right)$. Now, we want to further reduce it by fixing some of the quantities listed above. We take $\kappa=1/2$ and
\begin{equation}
    \beta_n=\frac{\lambda}{n!}\ ,\quad \beta_m=\frac{\lambda}{m!}\ .
\end{equation}
Now, the only free parameters are $\left(D,n,m,\lambda\right)$. Since we are working with a real scalar field, we must take $n>2$ and $m>2$ so that:
\begin{equation}
    \varphi=\left[-\frac{\left(n-1\right)!}{\left(m-1\right)!}\right]^{\frac{1}{n-m}}\in\R\ .
\end{equation}
This can be achieved by requiring:
\begin{equation}
    \frac{1}{n-m}\equiv k\in\Z\ .
\end{equation}
In order for $n$ to be integer, we must have
\begin{equation}
    n=m\pm1\ ,
\end{equation}
where $k=\pm 1$ is an integer parameter classifying alternative theories.
Since the $k=\pm 1$ theories are equivalent under exchange of $n$ and $m$, we can take, without loss of generality, $n=m+1$. Hence, we get
\begin{equation}
\begin{split}
\varphi&=-m\ ,\\
    R&=\frac{D}{D-2}\frac{\lambda}{2}\frac{\left(-m\right)^m}{\left(m+1\right)!}\ .
\end{split}
\end{equation}
In the above expressions, $m$ and $D$ are positive natural numbers, with $D>2$ for gravity to be dynamical, and $\lambda$ is a positive real number. One can straightforwardly observe that the moduli of $\varphi$ and $R$ grow with $m$, where $R$ is \textit{positive} for \textit{even} $m$ and \textit{negative} for \textit{odd} $m$. Namely, odd and even $m$ realise Anti de Sitter and de Sitter space-time, respectively. Given our assumptions on the theory parameters, the action reduces to:
\begin{equation}\label{reda}
    \mathcal{S}\left[g,\varphi\right]\equiv\int\diff^Dx\sqrt{g}\left[R+\frac{1}{2}\left(\nabla\varphi\right)^2+\frac{\lambda\varphi^{m}}{m!}+\frac{\lambda\varphi^{m+1}}{\left(m+1\right)!}\right]\ .
\end{equation}
This can be directly connected to the derivation of flow equations from an action performed in \eqref{gfa}. Namely, we can use \eqref{reda} to read off the flow equations
\begin{equation}\label{acteq}
    \begin{split}
        \frac{\de g_{\mu\nu}}{\de\lambda}=&-2R_{\mu\nu}-g_{\mu\nu}\sum_{n=1}^{+\infty}ns_{\ n}^{(D)}\varphi^{n-1}+4\frac{D-1}{D-2}\nabla_{\mu}\varphi\nabla_{\nu}\varphi-2\frac{5D-6}{D-2}\nabla_{\nu}\nabla_{\mu}\varphi+\\
        &-2\frac{D-1}{D-2}g_{\mu\nu}\nabla^{2}\varphi+2\frac{D-1}{D-2}g_{\mu\nu}\left(\nabla \varphi\right)^2+\mathcal{L}_{\xi}g_{\mu\nu}\ ,\\
        \frac{\de \varphi}{\de\lambda}=&-R-\frac{D}{2}\sum_{n=1}^{+\infty}ns_{\ n}^{(D)} \varphi^{n-1}+\frac{\left(D-1\right)\left(D+2\right)}{D-2}\left(\nabla \varphi\right)^{2}+\\
        &-\frac{D^{2}+4D-6}{D-2}\nabla^{2} \varphi+\mathcal{L}_{\xi} \varphi\ ,
    \end{split}
\end{equation}
where
\begin{equation}
        s_{\ k}^{(D)}\equiv\sum_{n=0}^{k}\frac{g_{n}}{\left(k-n\right)!n!}\left(\frac{2}{2-D}\right)^{k-n}\left(\frac{4D-6}{D-2}\right)^{n/2}\ .
\end{equation}
and:
\begin{equation}
    \left(g_1,\dots,g_{m-1},g_m,g_{m+1},g_{m+2},\dots\right)=\left(0,\dots,0,-\lambda,-\lambda,0,\dots\right)\ .
\end{equation}
In particular, given a positive integer $l>0$, we get:
\begin{equation}
    \begin{split}
        s_{\ m-l}^{(D)}=&0\ ,\\
        s_{\ m}^{(D)}=&-\frac{\lambda}{m!}\left(\frac{4D-6}{D-2}\right)^{m/2}\ ,\\
        s_{\ m+l}^{(D)}=&-\frac{\lambda}{m!l!}\left(\frac{2}{2-D}\right)^{l}\left(\frac{4D-6}{D-2}\right)^{m/2}+\\
        &\quad-\frac{\lambda}{(m+1)!(l-1)!}\left(\frac{2}{2-D}\right)^{l-1}\left(\frac{4D-6}{D-2}\right)^{(m+1)/2}\ .
    \end{split}
\end{equation}
Neglecting the diffeomorphisms, defining
\begin{equation}
    F\left(\varphi\right)\equiv\sum_{n=1}^{+\infty}ns_{\ n}^{(D)} \varphi^{n-1}
\end{equation}
and plugging-in the $\nabla_\mu\varphi=0$ condition, the flow equations reduce to:
\begin{equation}
        \begin{split}
        \frac{\de g_{\mu\nu}}{\de\lambda}=&-2R_{\mu\nu}-g_{\mu\nu}F\left(\varphi\right)\ ,\\
        \frac{\de \varphi}{\de\lambda}=&-R-\frac{D}{2}F\left(\varphi\right)\ .
    \end{split}
\end{equation}
The first equation, exploiting the fact that we work with an Einstein manifold, can be turned into a flow equation for $R$ as:
\begin{equation}
    \Dot{R}=2\frac{R^2}{D}+F\left(\varphi\right)R\ .
\end{equation}
Thus, in $D=4$, we have
\begin{equation}
            \begin{split}
        \Dot{R}=&\frac{R^2}{2}+F\left(\varphi\right)R\ ,\\
        \Dot{\varphi}=&-R-2F\left(\varphi\right)\ .
    \end{split}
\end{equation}
and:
\begin{equation}
    \begin{split}
        s_{m-l}=&0\ ,\quad s_{m}=-\frac{\lambda}{m!}\sqrt{5^{m}}\ ,\\
        s_{m+l}=&\frac{\left(-1\right)^{l}}{l!}\left[1-\frac{ l\sqrt{5}}{(m+1)}\right]s_{m}\ .
    \end{split}
\end{equation}
Defining the quantity
\begin{equation}
    \mathcal{A}\left(g,\varphi\right)\equiv R+2F\left(\varphi\right)\ ,
\end{equation}
the flow equations simplify as:
\begin{equation}
                \begin{split}
        \Dot{R}=&\frac{R}{2}\mathcal{A}\left(g,\varphi\right)\ ,\\
        \Dot{\varphi}=&-\mathcal{A}\left(g,\varphi\right)\ .
    \end{split}
\end{equation}
Writing things explicitly, we get
\begin{equation}
    \mathcal{A}\left(g,\varphi\right)=R-\lambda\varphi^{m-1}\sum_{l=0}^{\infty}c_{l}\varphi^l\ ,
\end{equation}
where we have introduced:
\begin{equation}
    c_l\equiv2\left(m+l\right)\frac{\left(-1\right)^l}{l!}\left(1-\frac{l\sqrt{5}}{m+1}\right)\frac{1}{m!}\sqrt{5^m}\ .
\end{equation}
First of all, we observe:
\begin{equation}
    \lim_{l\to\infty}c_l=0\ .
\end{equation}
Given the assumptions, the values of the Ricci curvature and the scalar are:
\begin{equation}
\varphi=-m\ ,\quad R=\lambda\frac{\left(-m\right)^m}{\left(m+1\right)!}\ .
\end{equation}
By expressing $\lambda$ in terms of $R$, we can finally write
\begin{equation}
     \mathcal{A}\left(g,\varphi\right)=R\left(1-\frac{15}{32}\varphi^{m-1}\sum_{l=0}^{\infty}c_{l}\varphi^l\right)\equiv R\cdot G\left(\varphi\right)\ ,
\end{equation}
and rephrase the flow equations as:
\begin{equation}
                \begin{split}
        \Dot{R}=&R^2\cdot G\left(\varphi\right)/2\ ,\\
        \Dot{\varphi}=&-R\cdot G\left(\varphi\right)\ .
    \end{split}
\end{equation}
\subsection{Fixed $m$}
At this point, we take $m=4$ and allow $\varphi$ to go off-shell along the flow. Namely, we drop its relation with $m$ except for the initial point. This way, we get:
\begin{equation}
    c_l\equiv\frac{\left(-1\right)^l}{l!}\left(1-\frac{l}{\sqrt{5}}\right)\frac{25\left(4+l\right)}{12}\ .
\end{equation}
As a starting point, we take $\lambda=1$.
    \begin{figure}[H]
    \centering
    \includegraphics[width=0.8\linewidth]{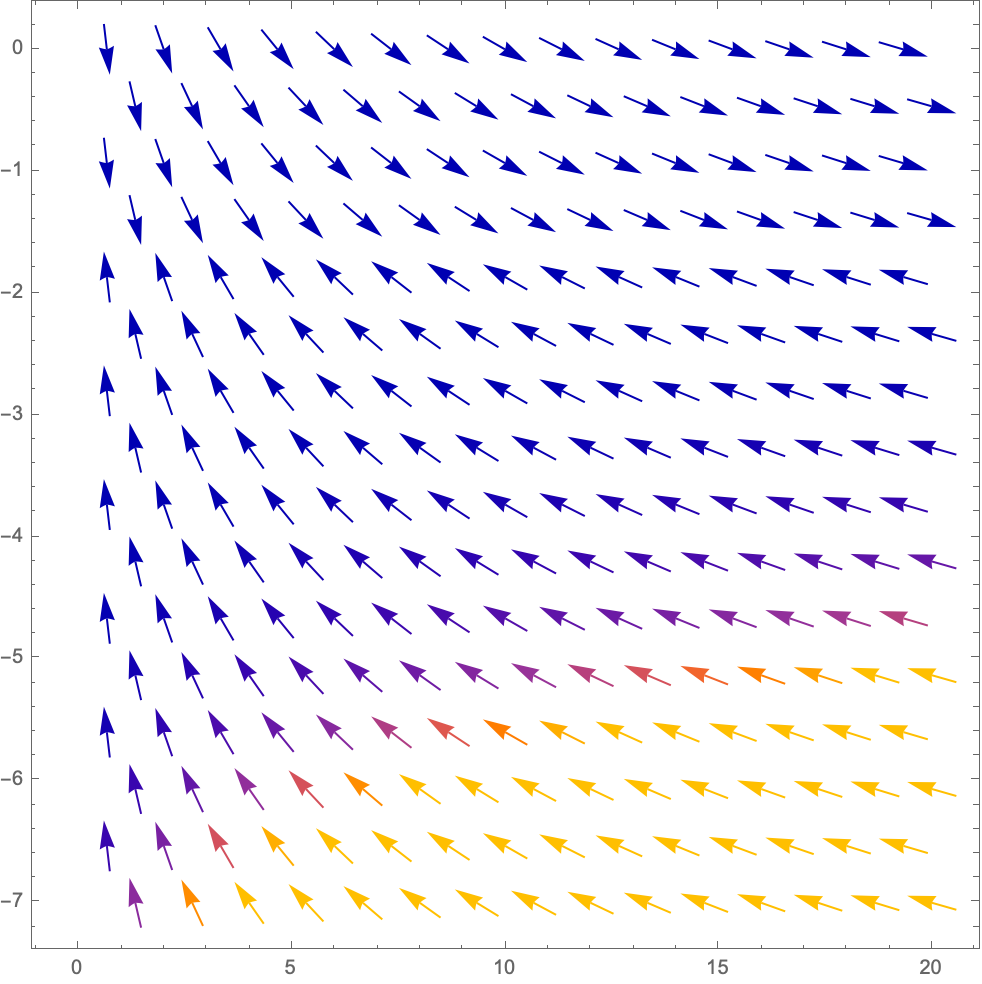}
    \caption{Plot of the vector field tangent to the flow, given generic initial conditions. In the above, the vertical axis corresponds to $\varphi$ while the horizontal one to $R$.}
\end{figure}
    \begin{figure}[H]
    \centering
    \includegraphics[width=0.8\linewidth]{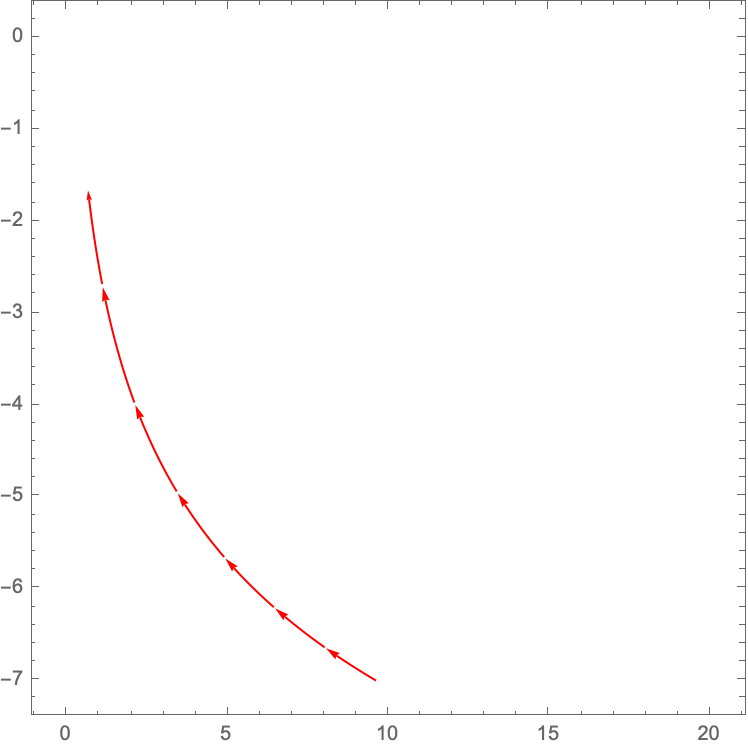}
    \caption{Plot of the vector field tangent to the flow, given the initial conditions $\varphi=-4$ and $R=32/15$. In the above, the vertical axis corresponds to $\varphi$ while the horizontal one to $R$.}
\end{figure}
As can be observed from the plot, starting from $\lambda=1$ and $m=4$, the system quickly goes to a fixed point.
\subsection{General analysis of the moduli space}
At this point, we take both $R$ and $\varphi$ to satisfy the on-shell conditions:
\begin{equation}
\varphi=-m\ ,\quad R=\lambda\frac{\left(-m\right)^m}{\left(m+1\right)!}\ .
\end{equation}
Being $R$ either positive (odd $m$) or negative (even $m$), the associated $4$-dimensional metric can always be written as
\begin{equation}
    \ds^2=\left(1-\frac{\Lambda r^2}{3}\right)\diff t^2+\left(1-\frac{\Lambda r^2}{3}\right)^{-1}\diff r^2+r^2\diff\Omega^2_2\ ,
\end{equation}
where the cosmological constant $\Lambda$ is:
\begin{equation}
    \Lambda=\frac{R}{4}=\frac{\lambda}{4}\frac{\left(-m\right)^m}{\left(m+1\right)!}\ .
\end{equation}
Now, we derive the entropy functional moving to the string frame version of the above action in $D=4$ as:
\begin{equation}
     \mathcal{F}=\int\diff^{D}x\sqrt{g}e^{-\varphi}\biggl[R+3\Delta \varphi+\left(\nabla \varphi\right)^2-\sum_{n=0}^{+\infty}s_{\ n}^{(D)}\varphi^{n}\biggr]\ .
\end{equation}
Plugging in the $\nabla_{\mu}\varphi=0$ condition and the expression for the metric and the scalar in terms of $m$ and $\lambda$, we have
\begin{equation}
\begin{split}
     \mathcal{F}\left[m,\lambda\right]&=-\int\diff^{4}x\ r^2\lambda\left(-m\right)^m e^{m}\biggl\{\frac{1}{\left(m+1\right)!}+\\
     &\quad+\frac{\sqrt{5^m}}{m!}\sum_{l=0}^{+\infty}\frac{m^{l}}{l!}\left[1-\frac{ l\sqrt{5}}{(m+1)}\right]\biggr\}=\\
     &\equiv\mathcal{T}\left(m,\lambda\right)\int\diff^{4}x\ r^2\ ,
\end{split}
\end{equation}
with:
\begin{equation}
    \mathcal{T}\left(m,\lambda\right)\equiv\lambda\left(-m\right)^m e^{m}\biggl\{\frac{1}{\left(m+1\right)!}+\frac{\sqrt{5^m}}{m!}\sum_{l=0}^{+\infty}\frac{m^{l}}{l!}\left[1-\frac{ l\sqrt{5}}{(m+1)}\right]\biggr\}\ .
\end{equation}
By inserting the above expression into the distance formula and assuming to start from a moduli space point with finite $\mathcal{T}$, we have:
\begin{equation}
    \Delta\sim|\log{\mathcal{T}\left(m,\lambda\right)}|\ .
\end{equation}
    \begin{figure}[H]
    \centering
    \includegraphics[width=0.8\linewidth]{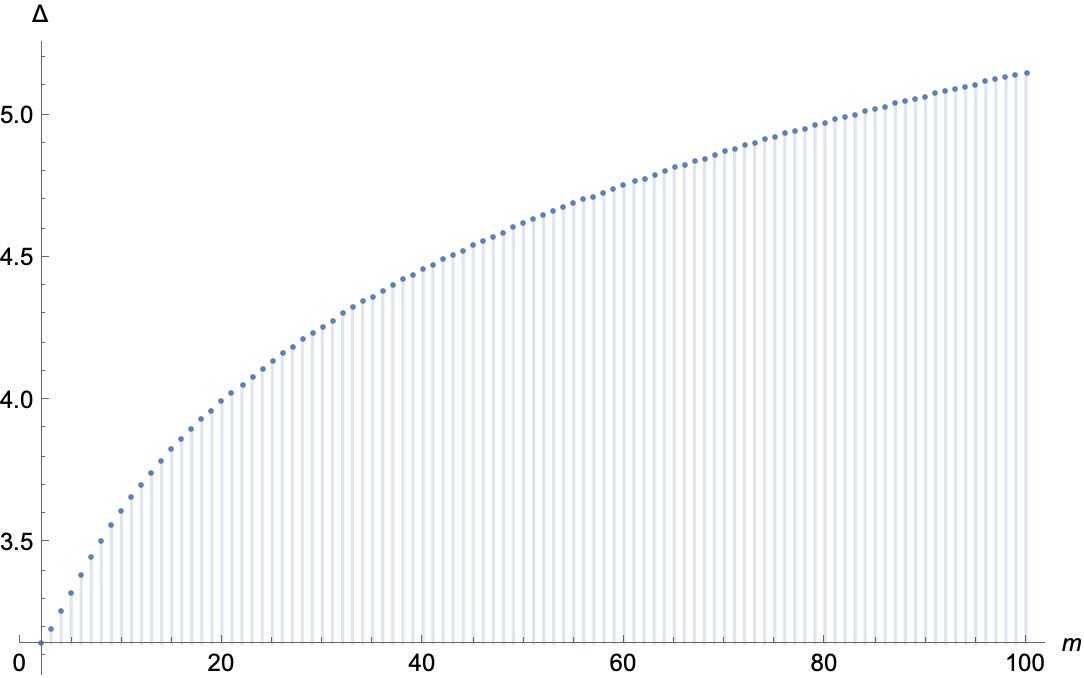}
    \caption{Behaviour of the distance $\Delta$ in $m$, when $\lambda$ is fixed to $1$.}
\end{figure}
In order to study the $m\to\infty$ limit, we observe that:
\begin{equation}
    \lim_{m\to\infty}|\mathcal{T}\left(m,\lambda\right)|\sim\lambda m^m e^{2m}\frac{\sqrt{5^m}}{m!}\rightarrow\infty\ .
\end{equation}
Thus, the $m\to\infty$ limit lies at infinite distance. Similarly, the $\lambda\to\infty$ limit trivially lies too at $\infty$ distance. How can the two limits be interpreted? Looking, once more, at the \textit{reduced} action
\begin{equation}
    \mathcal{S}\left[g,\varphi\right]\equiv\int\diff^Dx\sqrt{g}\left[R+\frac{1}{2}\left(\nabla\varphi\right)^2+\frac{\lambda\varphi^{m}}{m!}+\frac{\lambda\varphi^{m+1}}{\left(m+1\right)!}\right]
\end{equation}
in the usual frame, we can straightforwardly observe that:
\begin{itemize}
    \item The $\lambda\to\infty$ limit is the \textit{strongly} coupled regime of the theory, where amplitudes diverge and the perturbative picture breaks down.
    \item The $m\to\infty$ limit is the \textit{high-order} limit of the interactions, in which all the allowed scalar vertices one should consider in computing scattering amplitudes involve infinitely many legs. 
\end{itemize}
\newpage
\section{Conclusions}
In our work, a large class of flow equations for a metric-scalar system in $D$ dimensions was derived by performing volume-preserving variations of a general entropy functional $\mathcal{F}$ and considering its gradient flow. Thereafter, such formal discussion was connected to the physically grounded scenario in which the entropy functional emerges as the \textit{string frame} expression for a space-time action $\mathcal{S}$. The example of a massive scalar field with no higher-order self interactions in a cosmological constant background, which is arguably the first non trivial setting one might choose to consider, was analysed in detail. Indeed, the resulting flow equations turned out to be highly involved and produced an interesting flow behaviour for the simplest possible case: the one in which the scalar is constant in space-time and the metric is that of an Einstein manifold. Then, simple scalar bubble toy models where embedded in given space-time backgrounds. Their behaviour under \textit{Perelman's combined flow} was worked out in many different instances, leading to non-trivial fixed points and non-monotonic evolutions of the bubble radius. Thus, the flow behaviour of a \textit{metric} bubble in Anti de Sitter space-time was studied under \textit{normalised} flow equations for which any Einstein manifold is a fixed point, as a first step towards an attempt to connect considerations related to the \textit{Cobordism Conjecture} \cite{McNamara:2019rup} to the general framework of geometric flow equations. Thereafter, a simple solution of a metric scalar action $\mathcal{S}$ was forced to evolve precisely according to the flow equations derived from $\mathcal{S}$ in the way which was previously outlined, showing a non-trivial behaviour.
\newpage
\bibliographystyle{utphys}
\bibliography{bibliogra.bib}
\end{document}